\documentclass[11pt,a4paper]{iopart}

\pdfoutput=1
\usepackage{cleveref}
\usepackage{iopams}
\usepackage{harvard}
\usepackage{graphicx}
\DeclareGraphicsExtensions{.pdf,.eps,.png,.jpg}
\usepackage{subfigure}
\usepackage{color}
\usepackage{multirow}
\begin{document}

\title[Application of aritificial neural networks to CBC-GRB search]{Application of artificial neural network to search for gravitational-wave signals associated with short gamma-ray bursts}

\author{Kyungmin~Kim$^{1}$, Ian~W~Harry$^{2}$, Kari~A~Hodge$^{3}$, Young-Min~Kim$^{4}$, Chang-Hwan~Lee$^{4}$, Hyun~Kyu~Lee$^{1}$, John~J~Oh$^{5}$, Sang~Hoon~Oh$^{5}$, Edwin~J~Son$^{5}$}
\address{$^{1}$Department of Physics, Hanyang University, Seoul 133-791, Korea}

\address{$^{2}$Department of Physics, Syracuse University, Syracuse, NY 13244, U.S.A.}

\address{$^{3}$Department of Physics, California Institute of Technology, Pasadena, CA 91125, U.S.A.}

\address{$^{4}$Department of Physics, Pusan National University, Busan 609-735, Korea}



\address{$^{5}$Division of Computational Sciences in Mathematics, National Institute for Mathematical Sciences, Daejeon 305-811, Korea}



\date{\today}

\begin{abstract}
We apply a machine learning algorithm, the artificial neural network, to the search for gravitational-wave signals associated with short gamma-ray bursts. The multi-dimensional samples consisting of data corresponding to the statistical and physical quantities from the coherent search pipeline are fed into the artificial neural network to distinguish simulated gravitational-wave signals from background noise artifacts. Our result shows that the data classification efficiency at a fixed false alarm probability is improved by the artificial neural network in comparison to the conventional detection statistic. Therefore, this algorithm increases the distance at which a gravitational-wave signal could be observed in coincidence with a gamma-ray burst. In order to demonstrate the performance, we also evaluate a few seconds of gravitational-wave data segment using the trained networks and obtain the false alarm probability. We suggest that the artificial neural network can be a complementary method to the conventional detection statistic for identifying gravitational-wave signals related to the short gamma-ray bursts.





\end{abstract}

\ead{kmkim82@hanyang.ac.kr}


\maketitle


\section{Introduction}
\label{sec:intro}

Gamma-ray bursts (GRBs) are the most energetic electromagnetic events among various observable astronomical phenomena in the Universe. GRBs are very frequent events, as we observe them more or less once a day in isotropic spatial distribution \cite{GRBIsotropicDist}. In general, all observed GRBs can be classified in two categories, long and short, by a characteristic duration time, $T_{90}$,\footnote{It is defined by the time taken to accumulate $90\%$ of the burst fluence starting at $5\%$ of fluence level.} of the gamma-ray flashes. If the $T_{90}$ of a GRB is longer or shorter than 2 seconds,$\footnote{This fiducial time has been determined from the statistical distribution of observed duration times for BATSE sample. Note that $T_{90}=0.8$ seconds is used to distinguish between collapsar and non-collapsar progenitors for \emph{Swift} sample \cite{Bromberg:GRBClassification}.}$ it is classified as a long or short GRB, respectively. The most plausible scenario for the progenitor of long GRBs is a stellar collapse of a massive star to (i) a black hole with a forming accretion disk or (ii) a highly magnetized neutron star \cite{Ott:2009}.
In contrast, for short GRBs, it is believed that the inspiral merging process of a compact binary system composed of at least one neutron star such as a binary neutron star (BNS) or a neutron star-black hole binary (NS-BH) is the most viable progenitor model \cite{Berger:2013,shibata:2007}. 

The progenitor of short GRBs has been considered as one of the most promising sources of gravitational-waves (GWs) \cite{Collaboration:S6CBClowmass} that can be detected by the ground-based GW detectors such as the Laser Interferometer Gravitational-wave Observatory (LIGO) \cite{Abbott:2007kv} in the U.S. and Virgo \cite{Acernese:2008b} in Italy. Indeed, the LIGO scientific collaboration and the Virgo collaboration have conducted searches for GW signals from compact binary coalescences (CBCs) related to short GRBs (or CBC-GRB searches in short) with the data taken from two recent joint runs \cite{Abadie:2010cbcgrb,Abadie:2012grb,Aasi:2014IPNgrb}. The data of the first joint search \cite{Abadie:2010cbcgrb} were taken from the fifth LIGO science run (S5) and the first Virgo science run (VSR1) and the data of the second search \cite{Abadie:2012grb} were taken from the sixth LIGO science run (S6) and the second and third Virgo science runs (VSR2 and VSR3). On the other hand, the third joint search \cite{Aasi:2014IPNgrb} used the data obtained from both joint runs, S5/VSR1 and S6/VSR2,3, together. To try to observe relevant GW signals, the search pipelines for those previous searches use a matched filtering method \cite{Owen:1998dk} with template waveforms \cite{hexabank}. From the matched filtering, the search pipeline finds events which are highly correlated to the template waveforms by calculating the signal-to-noise ratio (SNR). When events in more than one detecter have an SNR that exceeds a predetermined search threshold, the events are recorded as a \emph{trigger}.
A trigger is characterized by several statistical quantities, for example, the signal-to-noise ratios (SNRs) in individual detectors, combined SNR statistics, values of various signal consistency tests, and other variants of them which are used for other consistency tests (refer to Ref. \cite{Harry:2011} for details). Thus, a trigger is characterized by a multi-dimensional vector of values. 

{\begin{table*}[t!]
\renewcommand{\arraystretch}{1.3}
\caption{The observationally-obtained information of our test GRBs. RA and DEC indicate right ascension and declination, respectively. The duration time $T_{90}$ is given in seconds. The values of duration time and redshift are obtained from Ref. \cite{Berger:2013}.} 
\label{GRBs}
\begin{tabular}{ l  c  c  c  c  c }
	\multirow{2}{*}{GRB} & \multicolumn{5}{c}{Observation} \\
	\cline{2-6}
	{} & UTC Time & RA & DEC & Duration, $T_{90}$ (sec)& Redshift \\
	\hline
	070714B & 2007-07-14 T04:59:29 & $57.85^\circ$ & $28.29^\circ$ &  2.0 & 0.923 \\
        070923 & 2007-09-23 T19:15:23 & $184.62^\circ$ & $-38.29^\circ$ & 0.05 & N/A \\
\end{tabular}
\end{table*}}

{\begin{table*}[ht!]
\renewcommand{\arraystretch}{1.3}
\caption{The characteristics of GW detectors related to selected GRBs in Table \ref{GRBs}. From the second to the fourth columns we summarize the antenna factors \cite{Anderson:2000yy,Anderson:2001} of each detector for `$+$' and `$\times$' polarizations of expected GWs  with pairing them in parentheses such as ($F_+$, $F_\times$). We present each detector's antenna response, ${\cal{F}}$, which is defined by ${\cal F}=(F_{+}^2 + F_{\times}^2)^{1/2}$, in the last three columns, respectively. 
} 
\label{GRBs_IFO}
\begin{center}
\begin{tabular}{ l  c  c  c  c  c  c  c}
	\multirow{2}{*}{GRB} & \multicolumn{3}{c}{Antenna Factor, $F_{+,\times}$} & \multicolumn{4}{c}{Antenna Response, ${\cal{F}}$} \\
	\cline{2-4}
	\cline{5-8}
	{} & H1 & L1 & V1 & H1 & {} & L1 & V1 \\
	\hline
	070714B & (-0.25, -0.07) & (0.27, 0.26) & (-0.83, -0.03) & 0.26 & {} & 0.37 & 0.83 \\
	070923 & (0.15, -0.28) & (-0.13, 0.37) & (0.56, 0.40) & 0.32 & {} & 0.39 & 0.69 \\
\end{tabular}
\end{center}
\end{table*}}

In the recent CBC-GRB searches, two different ranking methods, a likelihood ratio \cite{Abadie:2010cbcgrb} and a detection statistic \cite{Harry:2011} were calculated to classify whether a trigger is more likely to be caused by either a real GW signal radiated from an expected astrophysical source or a noise artifact originated by non-Gaussian and non-stationary noises coming from instruments and/or environments. 
The classification was done by the estimation of the false alarm probability (FAP) of a trigger based on the value of its detection statistic. The FAP of the triggers related to short GRBs were consistent with background; no evidence of GW signals in the GW data was found related to the considered short GRBs.

Over recent decades, various machine learning algorithms (MLAs) such as artificial neural networks (ANN) \cite{Hastie:2009,HechtNielsen:1989}, random forests of bagged decision trees \cite{Breiman:1996,Breiman:2001}, and support vector machines \cite{CortesV:1995,cristianini2000introduction} have been developed and evolved to analyze multi-dimensional data efficiently. Application of MLAs to many problems including several GW related searches \cite{Cannon:2008cqg,Biswas:2013prd,Adams:2013prd,Rampone:2013} have shown good classification performances for their nonlinear multi-dimensional data, which provides a complementary way of making a decision. Specifically, a recent development of an advanced ANN algorithms called \emph{Deep Learning} \cite{DengYu:2014} offers a clear motivation of applying ANN to GW data analysis. However, for the data set with a small sample size, relatively simple network structure can provide a satisfactory performance. In this motivation, we investigate the feasibility of applying ANNs to the CBC-GRB search as a new way of ranking a trigger. Moreover, we explore the possibility of the ANN as a potential method of improving the classification efficiency. Even without the notion of \emph{Deep Learning}, it is quite significant to apply a standard ANN to CBC-GRB classification for improving its detection performance, comparing to the convention detection statistic.


This paper is organized as follows. In Sec. \ref{DataPrep}, we summarize the data preparation. In Sec. \ref{performance}, we introduce the basics of the ANN employed in this work, summarize the methodology of our investigation, and discuss the classification performance test. In Sec. \ref{applications}, we present the results of our application of ANNs in terms of the detection sensitivity as a function of distance and the evaluation of unknown triggers. Finally, we summarize the results and discuss the future prospects of this approach in the era of the Advanced LIGO\footnote{https://www.advancedligo.mit.edu} and Advanced Virgo\footnote{https://wwwcascina.virgo.infn.it/advirgo} detectors in Sec. \ref{summary}.

\section{Data Preparation}
\label{DataPrep}

\subsection{Gravitational-Wave Data}

We focus on data from the fifth LIGO science run (S5) and the first Virgo science run (VSR1). The S5 data has been taken from the two LIGO detectors (H1 and H2) with 4 km and 2 km arms, respectively, in Hanford, Washington and a LIGO detector (L1) with 4 km arms, in Livingston, Louisiana. The VSR1 data has been taken from the 3 km arm Virgo detector (V1) at Cascina, Italy.  Among the 22 short GRBs observed when at least two of the LIGO and Virgo instruments were operating during the first joint search (S5/VSR1), we select two GRBs, GRB070714B \cite{Berger:2013,gcn6620,gcn6623,gcn6836} and GRB070923 \cite{Berger:2013,gcn6818,gcn6821}, which were observed by \emph{Swift} satellite \cite{Gehrels:2004} and have corresponding GW data in three detectors, H1, L1, and V1 available, coincidentally. The availability of using GW data is determined by the requirements of both stable operation at the event time of a GRB and provision of sufficient data (40 minutes in minimum) for the estimation of background distribution. We summarise the observation information of GRB070714B and GRB070923, including the event time and the sky location, in Table \ref{GRBs}, and the characteristics of corresponding GW detectors in Table \ref{GRBs_IFO}.

In the previous search \cite{Abadie:2010cbcgrb}, the authors concluded that they could find no evidence of GW signals related to these GRBs, GRB070714B and GRB070923; the estimated false alarm probabilities (FAPs) for the selected GRBs were consistent with the noise hypothesis. The exclusion distances to the potential progenitors of a GRB with 90\% confidence level were computed --- 3.2 Mpc to a BNS progenitor and 5.1 Mpc to a NS-BH progenitor for GRB070714B, and 5.1 Mpc to a BNS progenitor and 7.9 Mpc to a NS-BH progenitor for GRB070923.

{\begin{table*}[t!]
\renewcommand{\arraystretch}{1.1}
\caption{Brief description of the input variables we consider. One can find more detailed descriptions and forms of listed features in Ref. \cite{Harry:2011}.} \label{features}
\begin{tabular}{ l  l }
	Input Variable & Description \\
	\hline
	Single detector's SNR, $\rho^{}_{\textrm{IFO}}$ & Signal-to-noise ratio (SNR) value obtained from each \\
	{} & of the detectors' data where IFO = H1, L1, or V1. \\
	{} & In this work, we have $\rho^{}_\textrm{H1}$, $\rho^{}_\textrm{L1}$, and $\rho^{}_\textrm{V1}$ \\
	Coherent SNR, $\rho_{\textrm{coh}}$ & Coherent combination of the complex single detector's \\
	{} & SNRs \\
	Coherent $\chi^2$-test value & Mitigating non-Gaussian noise contribution by testing \\
	{} & the differences between template waveforms and instru-\\
	{} & mental/environmental noise \\
	New SNR, $\rho_{\textrm{new}}$ & Filtered $\rho_{\textrm{coh}}$ by checking whether the $\chi^2$-test value is \\
	{} & larger or smaller than  the number of degrees-of-freedom \\
	{} & of $\chi^2$ statistic \\
	Coherent bank $\chi^2$-test value & Testing the consistency of the observed $\rho_{\textrm{coh}}$ over \\
	{} & different template waveforms in the template bank at \\
	{} & the time of signal candidate trigger \\
	Coherent auto-correlation & Testing the consistency of the observed $\rho_{\textrm{coh}}$ over SNR \\
	$\chi^2$-test value & time series around the trigger in the template bank at \\
	{} & the time of signal candidate trigger \\
	Masses & Component masses of a binary system, $m_1$ (BH) and \\
	{} & $m_2$ (NS)\\
\end{tabular}
\end{table*}}

With the given GRBs' observation information, we run the coherent search pipeline \cite{Harry:2011}, which has been used for a recent CBC-GRB search \cite{Abadie:2012grb}, on the data from the sixth LIGO science run (S6) and the second and third Virgo science runs (VSR2 and VSR3). As a first step, the pipeline coherently combines the data from the three operational detectors, H1, L1, and V1.
Next, the pipeline divides the combined data into several partial segments such as \emph{the on-source} and \emph{off-source} segments (see Refs. \cite{Abadie:2010cbcgrb} and \cite{Abadie:2012grb} for details of the segmentation). The on-source segment is [-5,+1) seconds around the event times of selected GRBs. The off-source segment is 1944 seconds of data around the on-source segment that does not overlap with the on-source segment. 

The coherent search pipeline performs \emph{matched filtering} \cite{Owen:1998dk} on the data with \emph{the template bank of GW waveforms} \cite{hexabank}. 
The products of matched filtering are called as \textit{triggers}: the on-source triggers denote triggers that may contain a potential GW event candidate and the off-source triggers are believed to be noise artifacts originated by instrumental and/or environmental noises. The off-source triggers are used for the background estimation of the on-source triggers.
These on- and off-source triggers are characterized by several statistical quantities and physical parameters such as signal-to-noise ratio (SNR), various signal consistency tests, and component masses of a binary system. The detection statistic \cite{Harry:2011} of the coherent search pipeline can be calculated with these quantities and the estimation of FAP becomes possible. Then, we determine whether an on-source trigger is a GW signal or not based on the estimated FAP.

The coherent search pipeline also adds simulated waveforms\footnote{The template waveforms also can be called as simulated waveforms. However, the template waveforms are used only for the matched filtering. Thus, we use the term of simulated waveform only for the software injection to prevent confusion.} to the off-source segment\footnote{The on-source segment is not used in order to avoid any contaminations in the potential candidate GW event in the on-source segment.} and this procedure is referred to \emph{software injections}. Then the search pipeline repeats the matched filtering on the software injected waveforms.
The simulated waveforms for the software injection are generated by the spinning TaylorT4 waveform \cite{Boyle:2007TaylorT4} code in the LIGO Algorithm Library (LAL)\footnote{https://www.lsc-group.phys.uwm.edu/daswg/projects/lal/nightly/docs/html/} with correction terms up to the 3.5 post-Newtonian order. In the generation of those simulated waveforms, several physical parameters such as the masses and spins of component objects (NS or BH), and the distance to the expected binary system need to be chosen. The details of choosing physical parameters and their distributions are discussed in Appendix \ref{app_SoftInjParams}. From this procedure, we get software injection triggers which are separated into two categories. 
If there is a trigger found within in 100 ms of the time of simulation, we call that trigger as a \emph{found injection trigger} \cite{Predoi:thesis}. While, if not, it becomes a \emph{missed injection trigger}.



{
\begin{figure*}[ht!]
\centering
	\subfigure[NS-BH model]{\label{scatter_nsbh}\includegraphics[width=0.45\textwidth]{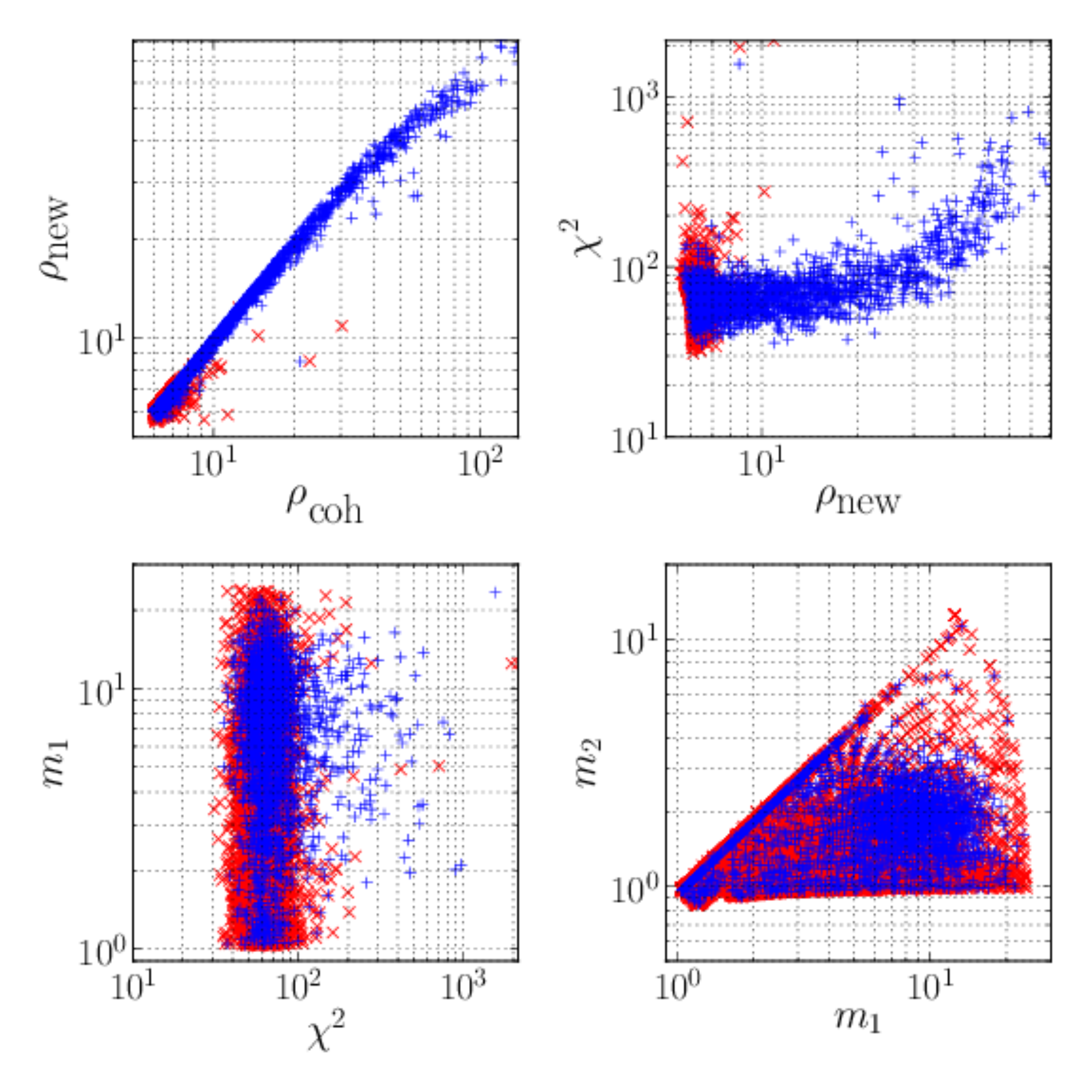}}
	\subfigure[BNS model]{\label{scatter_bns}\includegraphics[width=0.45\textwidth]{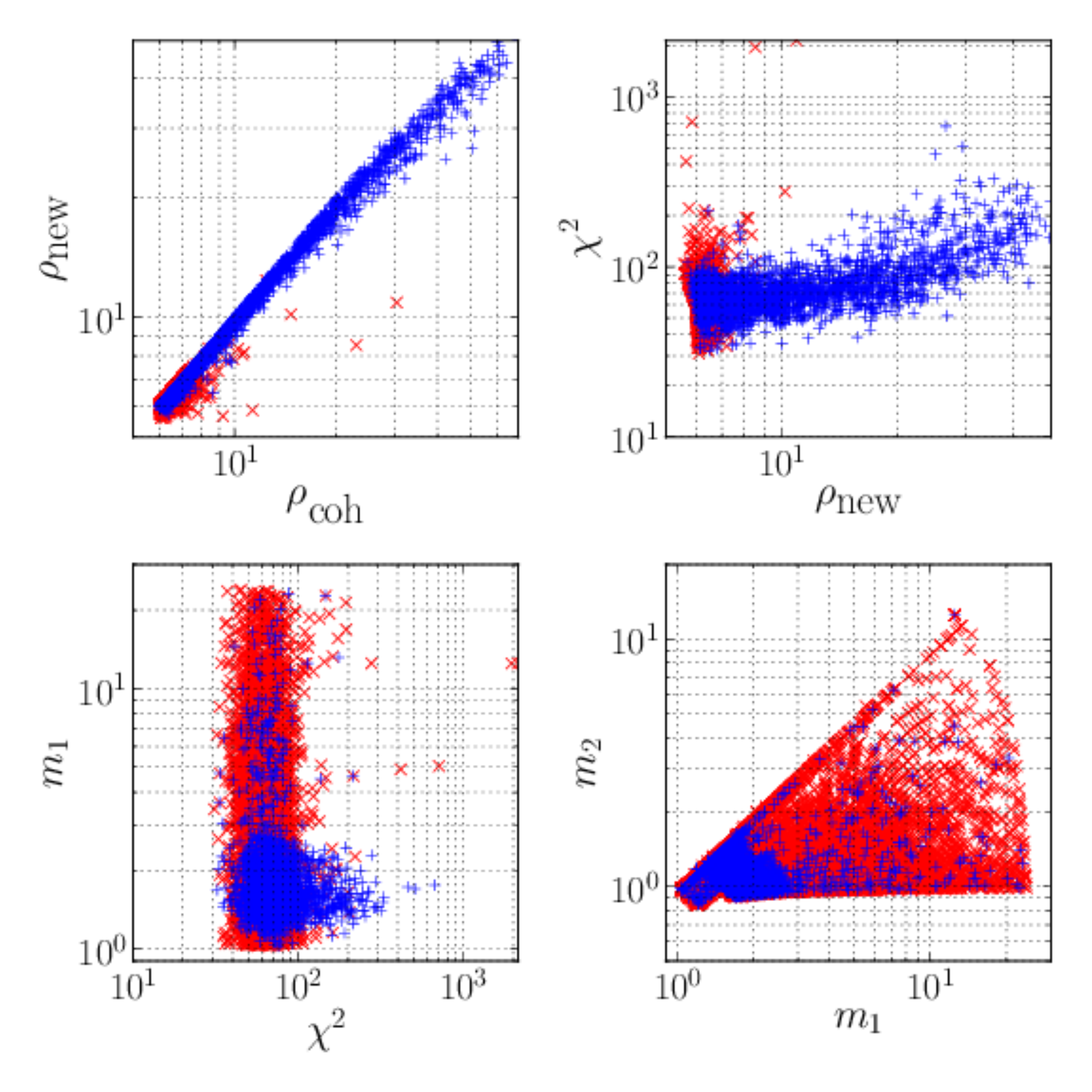}}
\caption{(Color online) The scatter plots of selected input variables for GRB070714B. In each figure, a blue $+$ denotes the signal sample and a red `$\times$' denotes the background sample. As one can see, NS-BH and BNS model cases have the same distribution in background samples. For signal sample, both models show similar distributions. However, the bottom panels show that mass related distributions are significantly different for both models.} 
\label{plot_features_scatter}
\end{figure*}}


\subsection{Training and Evaluation Samples}

We configure two different sample sets, training and evaluation samples for the ANN. For the performance test, we take the found injection triggers and off-source triggers as the sets of \emph{signal samples}, $X_S$, and \emph{background samples}, $X_B$, respectively. With this notation, we define similar notation $x^S$ and $x^B$ to denote each sample of $X_S$ and $X_B$, respectively, such as
\begin{eqnarray}
X_S^{} &=& \{x_l^S; l=1,2,\ldots,N_S^{}\}, \label{eq_sig_samples}\\
X_B^{} &=& \{x_m^B; m=1,2,\ldots,N_B^{}\}, \label{eq_bg_samples}
\end{eqnarray}
where $N^{}_S$ and $N^{}_B$ correspond the total number of the signal samples and the background samples, respectively.
In the preparation of samples for the performance test, the missed injection triggers are discarded because they are improper for the purpose of training ANN. Meanwhile, the on-source triggers are also not used for the performance test. However, we use the on-source triggers as another set of evaluation samples in order to find the most significant (or, equivalently, the loudest) trigger among them and to examine its significance.

\subsection{Input Variables}

For the configuration of the input variables of the multi-dimensional sample data, we take the 8 statistical and 2 physical quantities from the CBC-GRB search pipeline as input variables, since they are considered important in discriminating signals from noise and in characterizing the GW sources\footnote{We investigated  the influence of change in input variables by reducing the number of input variables using linear correlation coefficient \cite{Pearson:1895} and Principal Component Analysis (PCA) \cite{PCA:2008}. Reduction of input vector size did not significantly either improve or lower the classification performance. Thus we decided to use every possible variable in order to maximize the mount of information fed into ANN unless it impaired the performance.}. We choose three single detectors' SNRs ($\rho^{}_\textrm{H1}$, $\rho^{}_\textrm{L1}$, and $\rho^{}_\textrm{V1}$), coherent SNR ($\rho_{\textrm{coh}}$), coherent $\chi^2_{}$-test value, new SNR ($\rho_{\textrm{new}}$) of the coincidence, coherent bank $\chi^2_{}$-test value, coherent auto-correlation $\chi^2$-test value, and component masses, $m_1$ and $m_2$ for each trigger. The selected input variables are tabulated in Table \ref{features} with brief descriptions (For more details, see Ref. \cite{Harry:2011}). We plot the scatter plots of several input variables, the coherent SNR, the new SNR, the $\chi^2_{}$-test value,  and the component masses ($m_1$ and $m_2$) in Fig. \ref{plot_features_scatter}. From the figure, one can see that the distributions of statistical quantities have similar shapes between the NS-BH and the BNS model. The mass-related distributions for the signal samples are different, as expected given the injected ranges on masses for NS-BH versus BNS systems (Appendix \ref{app_SoftInjParams}). The background samples are exactly the same for both NS-BH and BNS models because we use the same bank of template waveforms for both models.

\section{Classification Performance Test using Artifical Neural Network}
\label{performance}

\subsection{Artificial Neural Network}
\label{MLAs}

The ANN \cite{Hastie:2009,HechtNielsen:1989} is a widely-used machine learning algorithm based on mimicking the biological neural system, which has been designed for artificial intelligence. This algorithm can be simplified with some mathematical models, which work as a black box system with data-driven input and output samples.

The implemented mathematical model includes nodes, a network topology, and learning rules adopted to a specific data processing task. Nodes are described by their number of inputs/outputs and the connection weights associated with each input and output. The network topology is closely related to the connections between the nodes. The learning rules represent how the connection weights are optimized. A node can be activated if the summed value of input nodes exceeds its threshold value.

\graphicspath{{figures/}}
\begin{figure}[t!]
\centering
\includegraphics[width=0.6\textwidth]{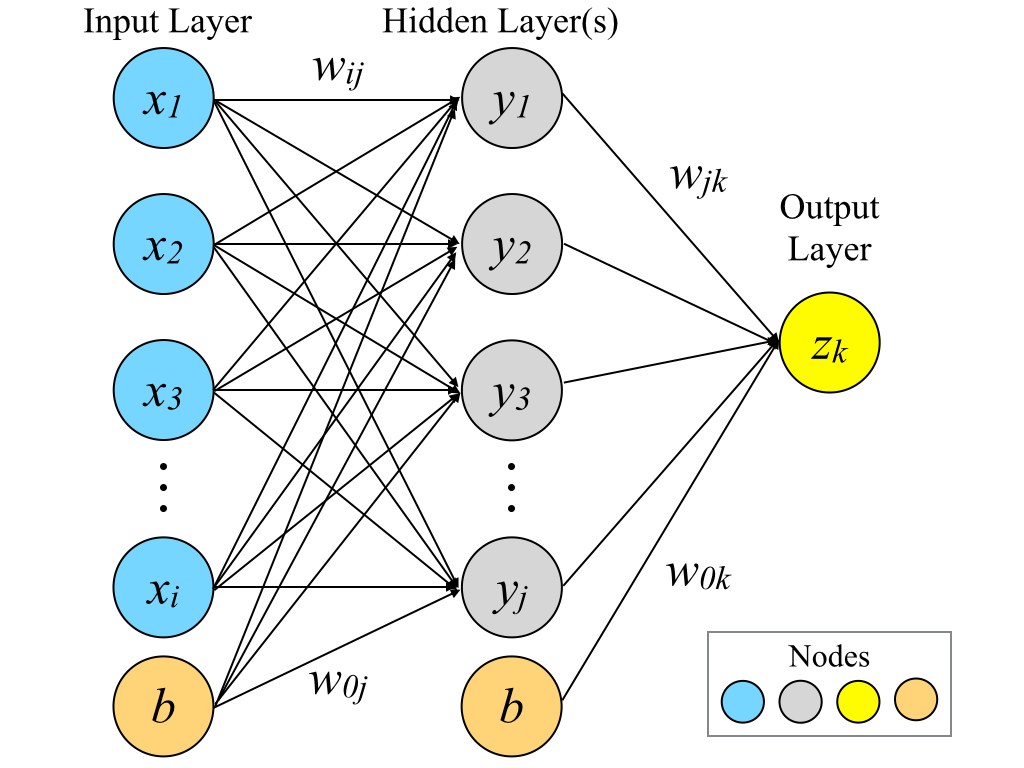}
\caption{(Color online) A schematic example of an ANN. In this example, the network topology consists of 1 input layer, 1 hidden layer, and 1 output layer. Each colored circle corresponds to a node. In this figure, $x_i$, $y_j$, and $z_k$ represent nodes and $w_{ij}$ and $w_{jk}$ represent connection weights. In specific, $b$s are the bias nodes and $w_{0j}$ and $w_{0k}$ are the connection weights from the bias nodes to the node(s) in next adjacent layers.} \label{ann_scheme}
\end{figure}

Among various models of ANNs with different topologies, {\it the multi-layered perceptron model} \cite{rosenblatt:1961} is widely used for its efficient classification. The model is composed of input and output layers as well as a few hidden layers in between. We present a simple schematic example of a network topology in Fig. \ref{ann_scheme}. 
For the nodes $x_i$ in the input layer, the features of an input sample are used. Those nodes in the input layer are connected to the nodes $y_j$ in the adjacent hidden layer with different connection weights $w_{ij}$. Then the value of $y_j$ is determined by an activation function $f$ as
\begin{equation}
y_j = f({\cal N}_j), \label{eq_x_j}
\end{equation}
and ${\cal N}_j$ is defined by a linear combination of the $x_i$ and $b$ such as
\begin{equation} 
{\cal N}_j \equiv \sum_{i=1}^i x_i \cdot w_{ij} + b \cdot w_{0j}, \label{eq_net}
\end{equation}
where $b$ denotes the bias node that makes the node $y_j$ activated and $w_{0j}$ indicates the connection weight between the bias node in the input layer and the node $y_j$. 
The activation function can be chosen in various options 
such as sigmoid, piecewise linear, step, and gaussian functions. Here, we choose the sigmoid function 
\begin{equation}
f({\cal N}_j) = (1 + e^{-2 s {\cal N}_j})^{-1}, \label{activation_func}
\end{equation}
which is the activation function chosen in Ref. \cite{Biswas:2013prd}. From Eq. (\ref{activation_func}), one can see that the property of this sigmoid function is determined by the steepness $s$.
Also, if many hidden layers are given in a network topology, similar processes are repeated until the connections will converge in the output layer.



In each layer, the learning algorithm finds the optimal connection weights between nodes. We particularly use the improved resilient back propagation (iRPROP) algorithm \cite{IgelHusk:2000}, which minimizes the error between the output and the goal values. We choose the Fast Artificial Neural Network (FANN) library package \cite{fann} for machine learning and find optimal connection weights by controlling parameters such as the number of layers, the number of nodes in each layer, and so on, which are given in the library. 

We construct a simple network topology with one input layer, one hidden layer, and one output layer. For our feature space of 10 input variables, we let 10 nodes be placed in the input layer and put the same number of nodes in the hidden layer. In the output layer, we have only 1 output node. Bias nodes in the input and the hidden layers are systematically placed with a value of 1.0 in our model. For the steepness $s$ of the sigmoid activation function, we simply set it to be $0.5$. These parameters are empirically determined by considering the computational expenses and the efficiency of the classification performance which will be described in Sec. \ref{performance_test_result}.

When the network topology and other training parameters are fixed, we can train the ANN with the training samples which have preassigned classes either 1 for a signal sample or 0 for a background sample. The goal of the training process is to find the most optimal set of connection weights such that  the error between the values assigned by the ANN and the target values of samples is minimized. In this work, the error is represented by the mean-squared-error (MSE) that is defined as
\begin{equation}
\textrm{MSE} \equiv \frac{1}{N} \sum_{k=1}^N \left| z_k^t - z_k^o\right|^2, \label{eq_mse}
\end{equation}
where $N$ is the total number of samples and $z^t_k$ and $z^o_k$ are the target value of the output and the observed value of a sample, respectively, of a sample. In this work, $z_k^t$ is the class (1 or 0) of a sample. Meanwhile, $z_k^o$ is determined by Eqs. (\ref{eq_x_j}) and (\ref{activation_func}) and $z_k$ has a value between 0 and 1. {\color{blue}\textbf{We terminate the training process when MSE stops decreasing and reaches a plateau with small oscillation around it as the iteration evolves.}}

Samples of known and unknown class can be evaluated deterministically by the trained ANN. The final $z_k$ value obtained after this evaluation process corresponds to the prediction of whether a sample is a signal or not. Hereafter, we call the final $z_k$ as the \emph{rank}, $r$, of a sample.

\subsection{Results}
\label{performance_test_result}

We introduce the receiver operating characteristic (ROC) curve in order to interpret the result. The ROC curve is obtained by calculating the efficiency and FAP which are defined as
\begin{eqnarray}
\textrm{Efficiency}(R) &\equiv& \frac{N^{}_S (R)}{N^{}_S}, \label{eq_eff} \\
\textrm{FAP}(R) &\equiv& \frac{N^{}_B (R)}{N_B}, \label{eq_fap} 
\end{eqnarray}
where $R$ denotes a threshold chosen among the ranks of evaluation samples. ${N}_S(R)$ and ${N}_B(R)$ in the numerators of Eqs. (\ref{eq_eff}) and (\ref{eq_fap}) are defined as 
\begin{eqnarray}
{N}_S(R) &\equiv& \{x_l^S(r);r \ge R, l = 1,2,\ldots,N_S\},\\
{N}_B(R) &\equiv& \{x_m^B(r);r \ge R, m = 1,2,\ldots,N_B\},
\end{eqnarray}
respectively, that is, the number of evaluated signal samples and background samples with scored ranks, $r$, exceeding a criterion rank value, $R$ among all $x_l^S(r)$ and $x_m^B(r)$.

When we compute Eqs. (\ref{eq_eff}) and (\ref{eq_fap}), we only have $\sim$ 2,000 - 3,000 signal samples and $\sim$ 6,000 - 7,000 background samples for the denominator $N_S$ and $N_B$, respectively. Therefore, if the numerator $N_B (R)$ is 1, the minimum value of FAP becomes $\sim$$10^{-4}$ due to the number of background samples. That value corresponds to a $\sim$$3.89$-$ \sigma$ confidence level in a sense of the normal distribution. However, since we desire a confidence level of at least $5$-$\sigma$ (FAP of $\sim$$10^{-7}$) to confidently claim the detection of a real GW signal event, using only the samples from a single segment limits our ability to determine significance on the expected level. In order to reach this minimum FAP, we need to include more background samples. This issue is not addressed in this paper, but we discuss how we can increase the background samples for future work in \ref{fap_lowering}.

{\graphicspath{{figures/}}
\begin{figure}[t!]
\centering
	\includegraphics[width=0.6\textwidth]{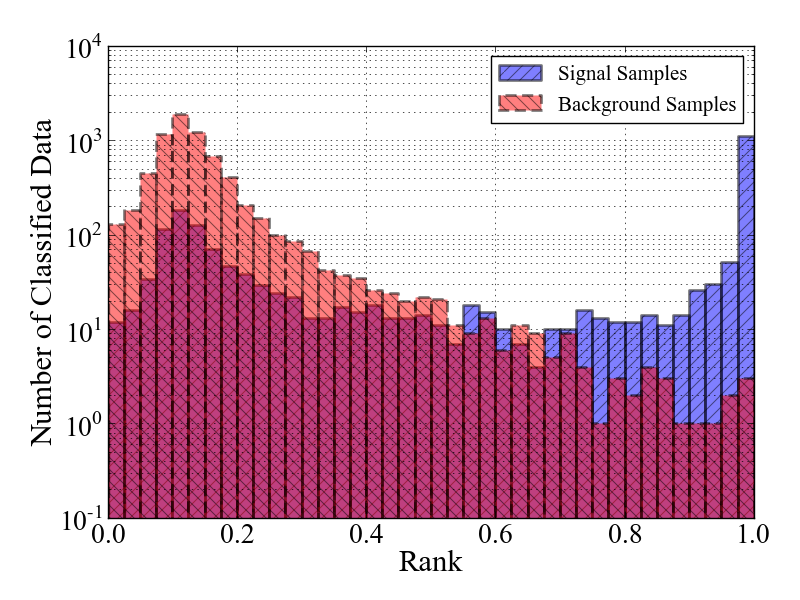}\caption{(Color online) Histogram of ranks scored on evaluation samples of NS-BH model for GRB070714B using the result of evaluation trial \#1.} \label{rank_hist_of_all_samples}
\end{figure}}

In order to fairly evaluate our classification efficiency, we split the full set of signal and background samples prepared in the previous section into training and evaluation samples via a \emph{round-robin process} (see \ref{round-robin} for details). Then we train the ANN with the training samples and test the classification performance with the evaluation samples. The training process ends with the MSEs reaching 0.08 and 0.06 for NS-BH model and BNS model, respectively.\footnote[1]{For this result, one can guess that it seems to be stuck in local minima. However, recent works \cite{Pascanu,Dauphin} explain that there does not exists distinct local minimum comparing to the global minimum in the higher dimensional non-convex optimization. Instead they are all similarly long plateaus or saddle points that leads to slow down the learning convergence.} We find that each of the trained ANNs result in different ranks on some of evaluation sample depending on the randomly given initial connection weights. Therefore, we repeat the training process 100 times in order to see distribution of rank and to obtain a representative statistical quantity for a given evaluation sample. The method for obtaining representative statistical quantity will be discussed later. From the repeated trials, it is shown that the resulted MSEs are similar to the values stated above for all trials of both NS-BH and BNS cases. Then we evaluate the evaluation samples with each trained classifier.
In Fig. \ref{rank_hist_of_all_samples}, we present a histogram of ranks scored on evaluation samples by using the result of evaluation trial \#1 out of 100 trials. One can see that the scored ranks of both signal and background samples are widely spreaded between 0 and 1. This result shows that the ANN does not classify the given data clearly. 


{\graphicspath{{figures/}}
\begin{figure}[t!]
\centering
	\includegraphics[width=0.6\textwidth]{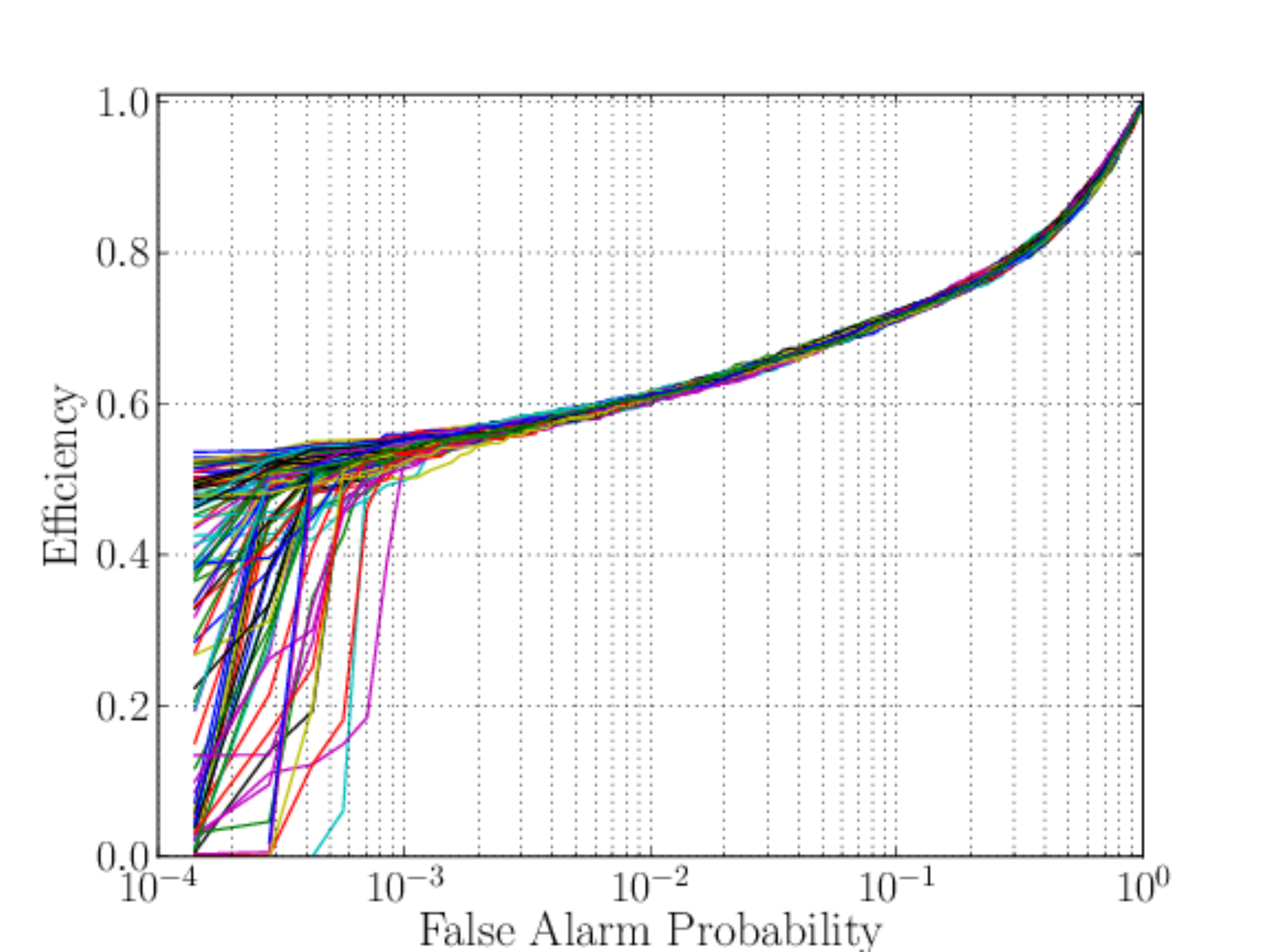}
\caption{(Color online) The ROC curves drawn via ranks of NS-BH model for GRB070714B. This figure is obtained from 100 trials of training and evaluation, using the same data and fixing the training parameters (except for the random seed) for all trials.} \label{roc_with_rank}
\end{figure}}

{\graphicspath{{figures/}}
\begin{figure}[t!]
\centering
	\includegraphics[width=0.6\textwidth]{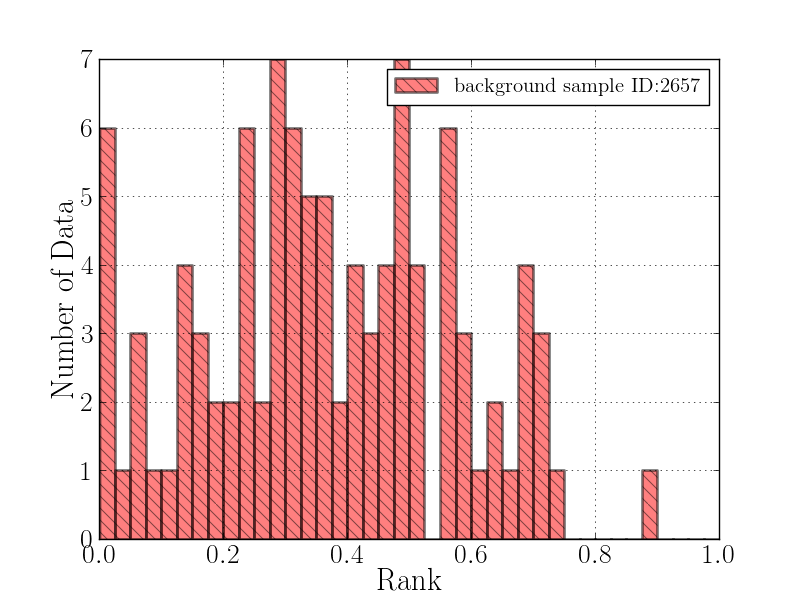}\caption{(Color online) The histogram of ranks scored on a background sample (ID \#2657) of NS-BH model for GRB070714B.} \label{rank_hist_of_a_sample}
\end{figure}}

{
\begin{figure*}[t!]
\centering
	\subfigure[NS-BH model for GRB070714B]{\label{roc_triple_nsns}\includegraphics[width=0.45\textwidth]{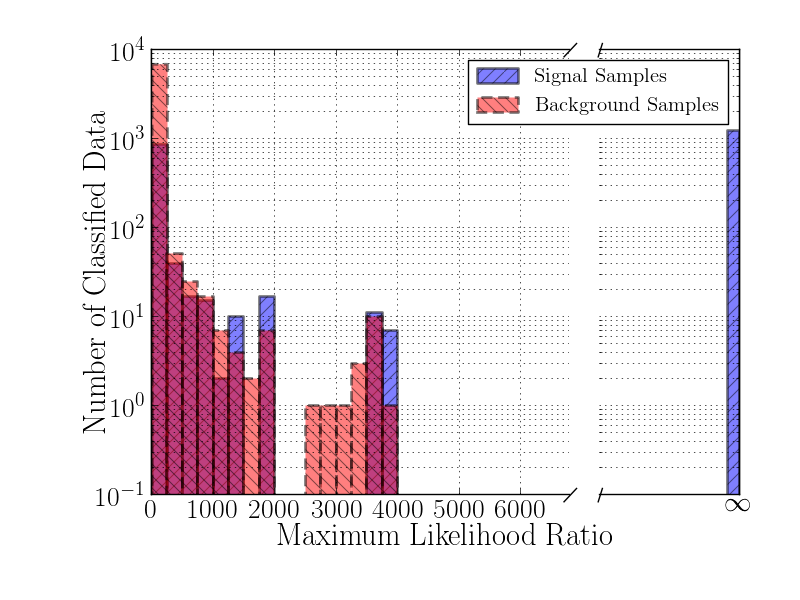}}
	\subfigure[BNS model for GRB070714B]{\label{roc_triple_nsns}\includegraphics[width=0.45\textwidth]{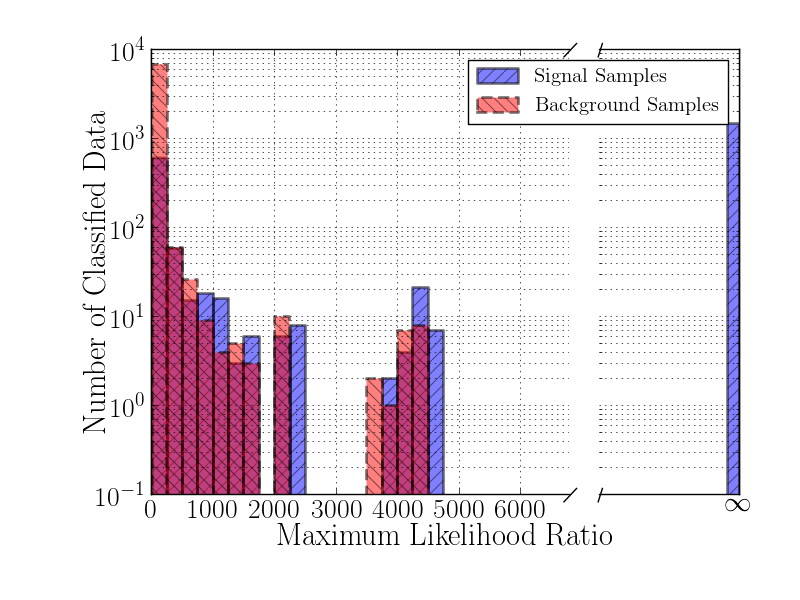}}
	\subfigure[NS-BH model for GRB070923]{\label{roc_triple_nsbh}\includegraphics[width=0.45\textwidth]{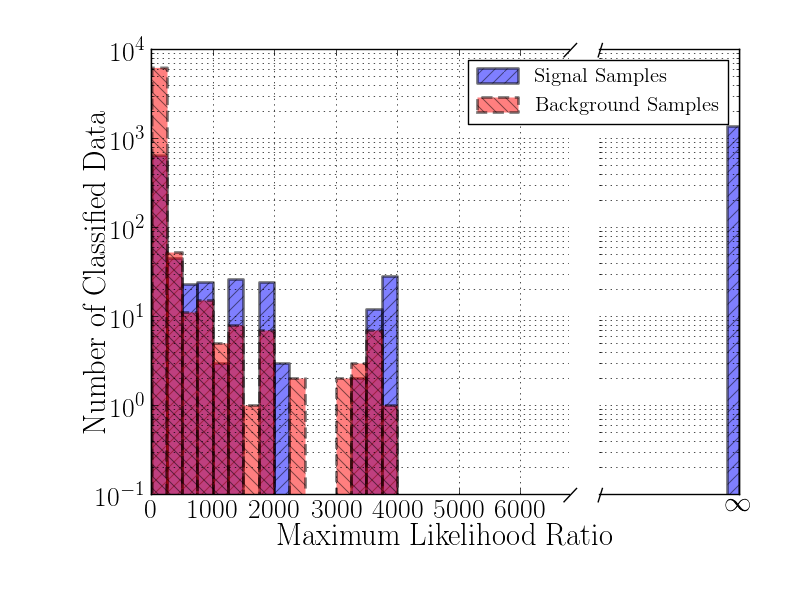}}
	\subfigure[BNS model for GRB070923]{\label{roc_triple_nsns}\includegraphics[width=0.45\textwidth]{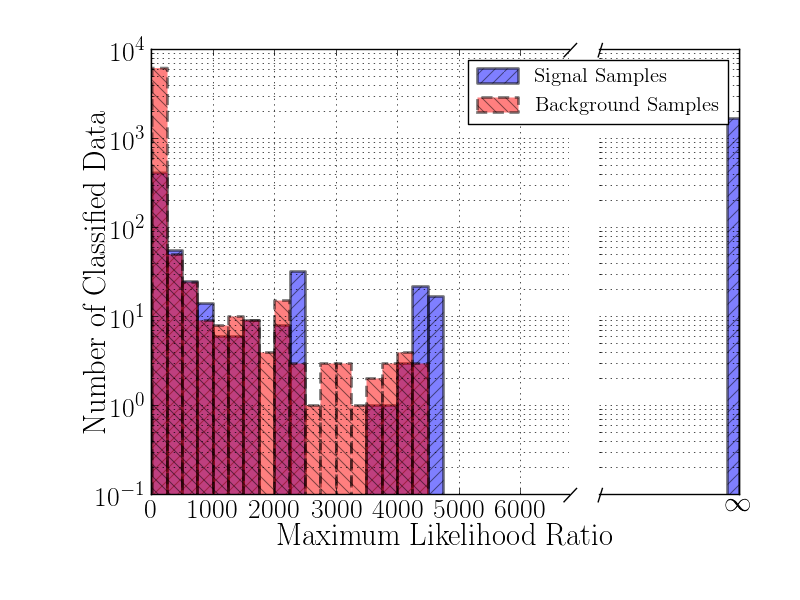}}
\caption{(Color online) The histogram of the maximum likelihood ratios of the signal and background samples. The blue bars and red bars denote the signal samples and the background samples, respectively. One can see that about half of signal samples are clearly separated from the distributions of background samples for all considered cases.} \label{fig_histogram_mlr}
\end{figure*}}

We also show the ROC curve of each trial in Fig. \ref{roc_with_rank}. The ROC curves are given in the Fig. \ref{roc_with_rank} by calculating Eqs. (\ref{eq_eff}) and (\ref{eq_fap}) by varying $R$ in a range from $R_{\textrm{min}}$ with $N_B(R_{\textrm{min}})=N_B$ to $R_{\textrm{max}}$ with $N_B(R_{\textrm{max}})=1$. In particular, if $N_B(R) = N_B$, Eq. (\ref{eq_fap}) becomes 1 and this special point corresponds to the point at the upper-right corner of the ROC curves. On the other hand, $N_B(R) = 1$ corresponds to the leftmost point, i.e., the minimum FAP of each curve. This variation in $R$ is consistently applied to both $N_S(R)$ and $N_B(R)$. 
As one can see from the ROC curves in Fig. \ref{roc_with_rank}, when we repeat the training and evaluation processes many times (100 trials in our case), we find that there are large variations in the efficiency below FAP of $\sim$ $10^{-3}$ of each run: We find that when (i) the maximum rank of all background samples is greater than that of all signal samples or (ii) the number of signal samples exceeding the maximum rank of background samples is small, the efficiency at the minimum FAP is significantly decreased. On the other hand, when (i) the maximum rank of background samples is smaller than the maximum rank of signal samples and (ii) there are many signal samples exceeding the maximum rank of background samples, the efficiency can be increased at the minimum FAP.

Simple statistics such as mean and standard deviation can hardly represent either the distribution of efficiencies at the minimum FAP or the distributions of ranks for the given samples as shown in Fig. \ref{rank_hist_of_a_sample}.
Therefore, in this work, we adopt the maximum likelihood ratio (MLR) which is shown to be an optimal method in obtaining a representative quantity for GW data \cite{Biswas:2012prd,Biswas:2013prd}.

{
\begin{figure*}[t!]
\centering
	\subfigure[NS-BH model for GRB070714B]{\label{roc_triple_nsbh}\includegraphics[width=0.45\textwidth]{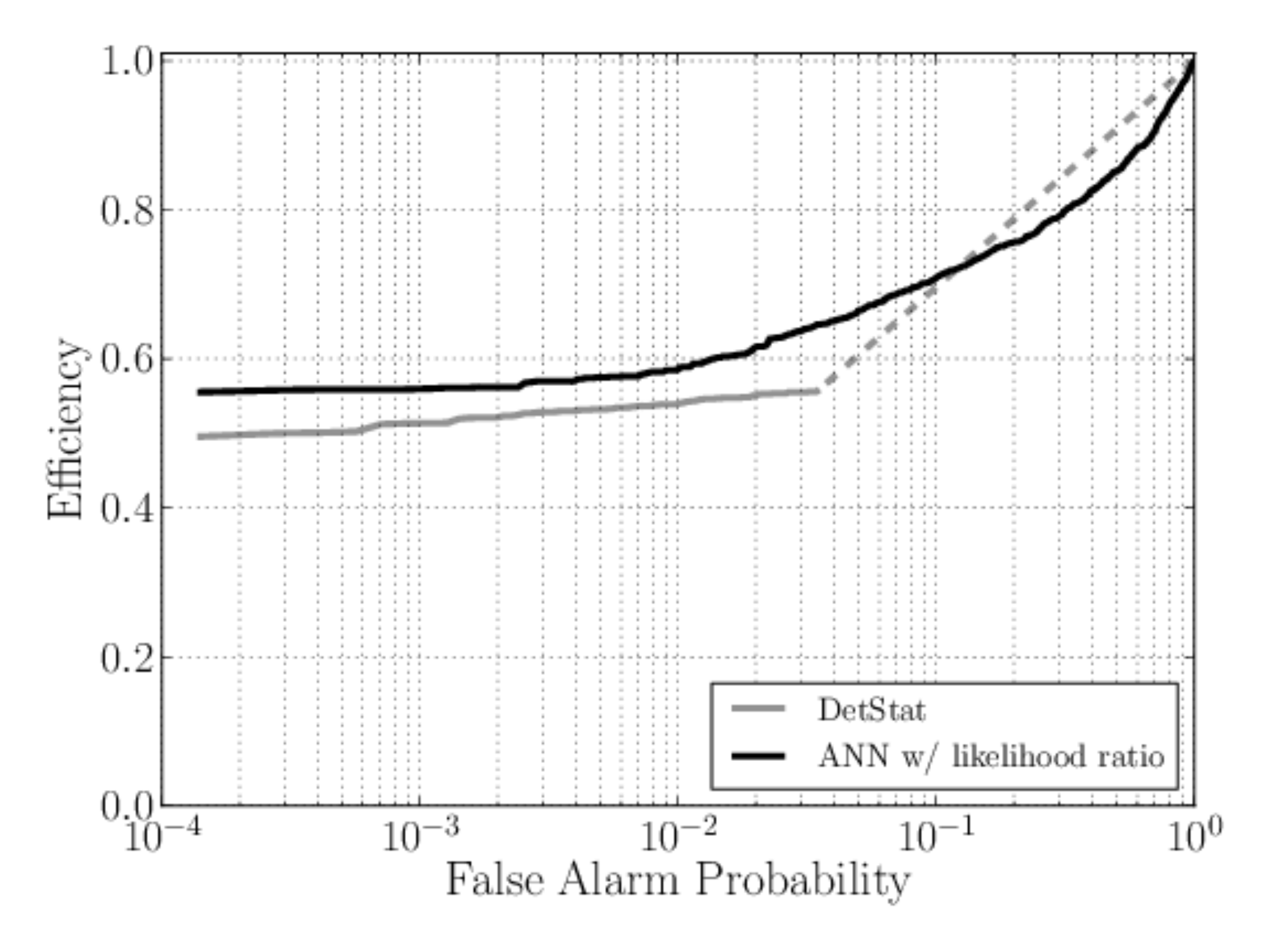}}
	\subfigure[BNS model for GRB070714B]{\label{roc_triple_nsns}\includegraphics[width=0.45\textwidth]{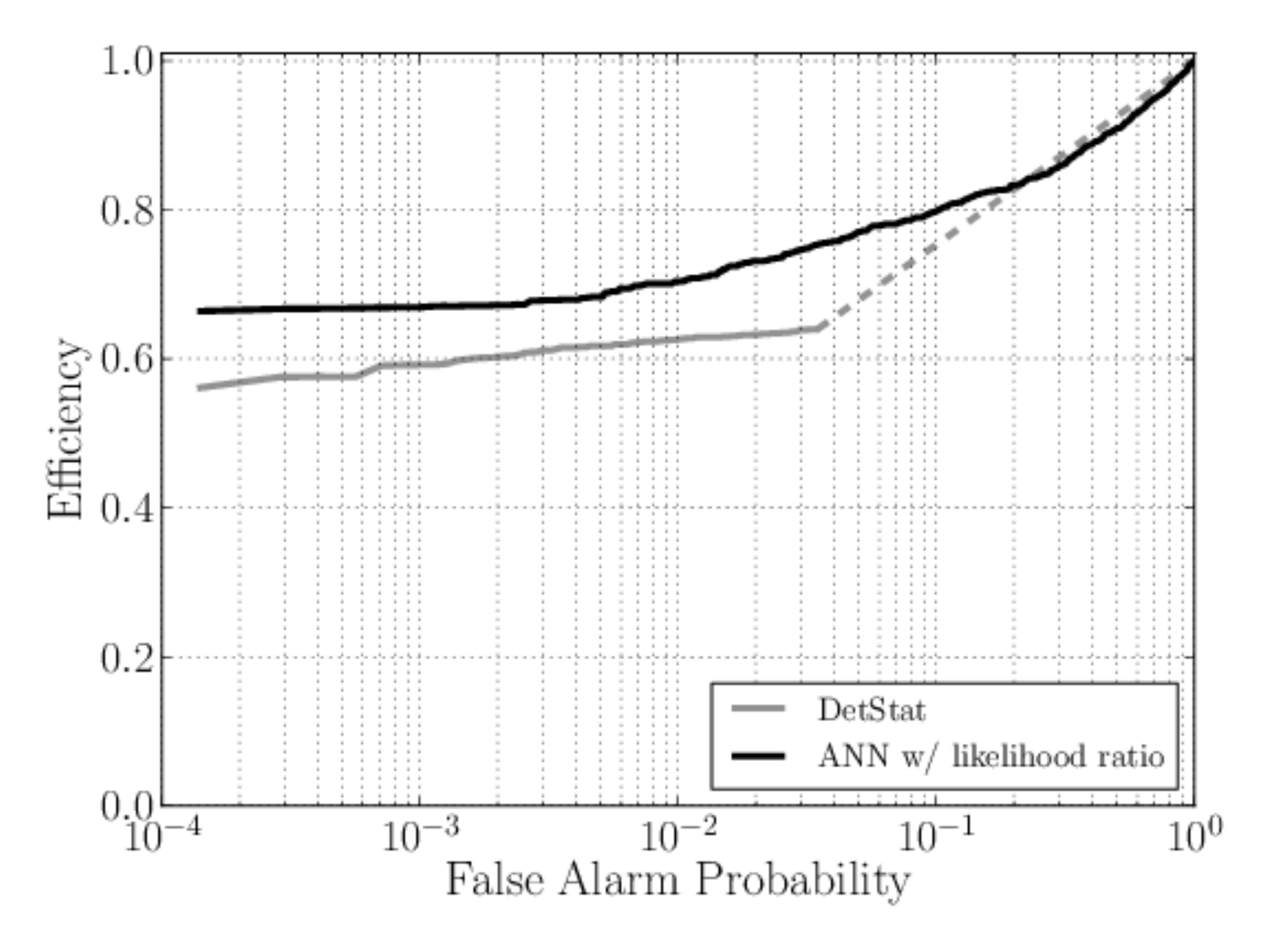}}
	\subfigure[NS-BH model for GRB070923]{\label{roc_triple_nsbh}\includegraphics[width=0.45\textwidth]{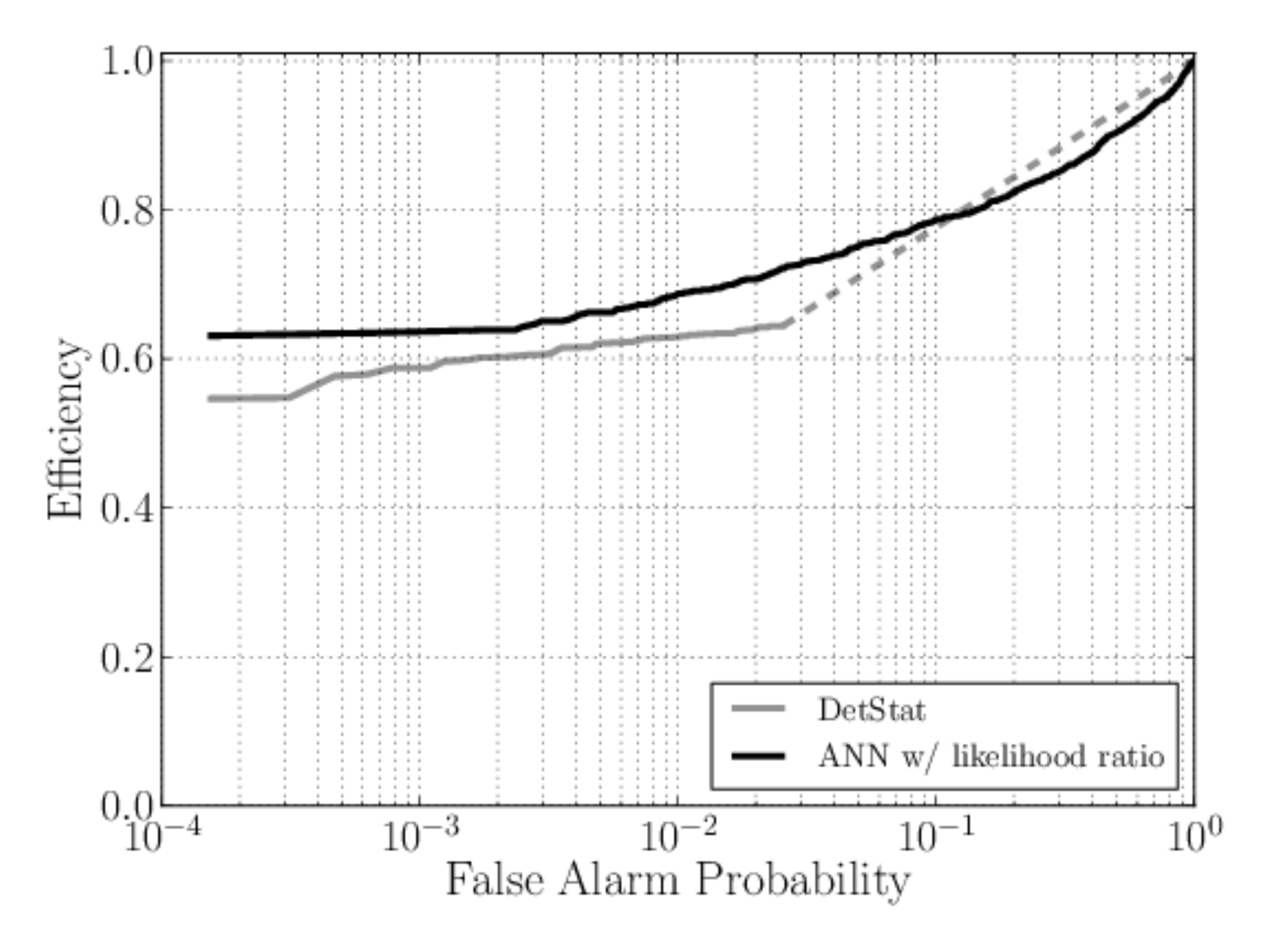}}
	\subfigure[BNS model for GRB070923]{\label{roc_triple_nsns}\includegraphics[width=0.45\textwidth]{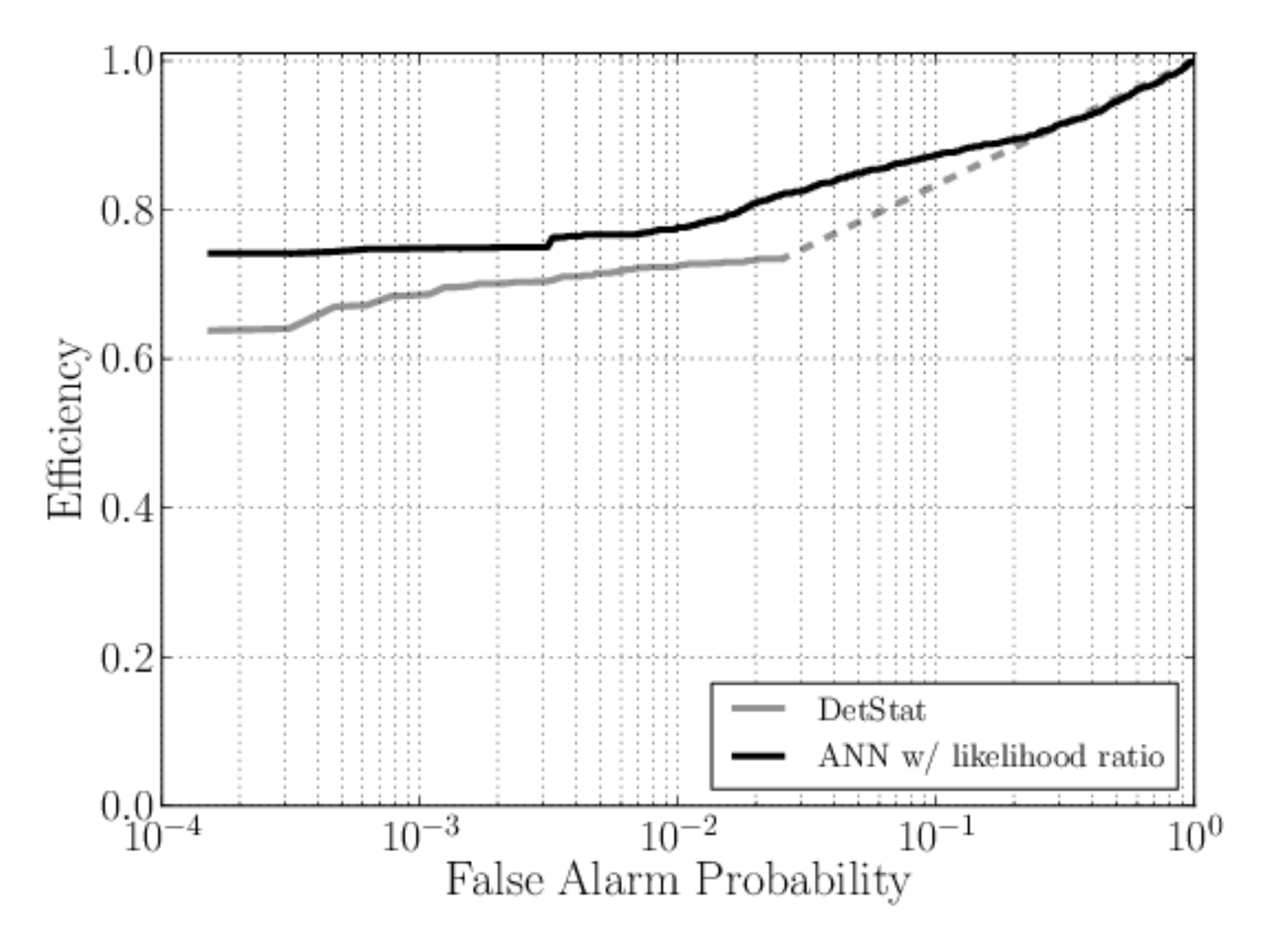}}
\caption{Comparion of the ROC curves of the ANNs' results against to the detection statistic of the coherent CBC-GRB search. The black solid line indicates the ROC curve of the ANNs with the MLR and the gray solid line indicates that of the detection statistic. The gray dashed-line is caused by the discontinuity of the calculated efficiency and FAP with the detection statistic.} \label{roccurve_mlr}
\end{figure*}}

The MLR for an $n$-th sample can be calculated by
\begin{equation}
\lambda(r_n) = \max_{r_n^\alpha} \left\{ \frac{\int^1_{r^{\alpha}_n} p(r^{\alpha '}_n | 1) dr^{\alpha '}_n}{\int^1_{r^{\alpha}_n} p(r^{\alpha '}_n | 0) dr^{\alpha '}_n} \right\} = \max_{r_n^\alpha} \left\{ \frac{P_1 (r_n^\alpha)}{P_0 (r_n^\alpha)} \right\}, \label{eq_max_likelihood_ratio}
\end{equation}
where $\alpha$ denotes the trial ($\alpha=1,2,\ldots,100$) and $r_i^{\alpha}$ indicates the rank of $i$-th sample in a trial $\alpha$. As one can see from Eq. (\ref{eq_max_likelihood_ratio}), we find the maximum value of the ratio between the probability of correctly
finding a true signal ($P_1$) and the probability of finding a false signal ($P_0$). Thus, in our case, Eq. (\ref{eq_max_likelihood_ratio}) can be rewritten as
\begin{equation}
\lambda(r_n) = \max_{r_n^\alpha} \left\{ \frac{ \textrm{Efficiency} (r_n^\alpha)}{ \textrm{FAP} (r_n^\alpha)} \right\}.
\end{equation}
In Fig. \ref{fig_histogram_mlr}, we plot the distribution of MLRs. From this figure, one can see that the background samples have finite values of MLR. However, for signal samples, about half of samples have similar values of MLR to the values of background samples and the rest of them are separated from the range of finite MLRs and take an infinite value. This tendency is consistently shown in all considered cases. The infinity MLR of signal sample can be easily derived from Eq. (\ref{eq_max_likelihood_ratio}), i.e., if there are no background samples exceeding $r_n$, then FAP($r_n$) becomes 0 and it leads $\lambda(r_n)$ to be infinity. Also, when we compare this Fig. \ref{fig_histogram_mlr} to the histogram of ranks in Fig. \ref{rank_hist_of_all_samples}, we see that the classification efficiency with the MLR is enhanced by the fraction of clearly separated signal samples.

With the MLR, the efficiency and FAP of Eqs. (\ref{eq_eff}) and (\ref{eq_fap}) are changed to
\begin{eqnarray}
\textrm{Efficiency}(\Lambda) &\equiv& \frac{N^{}_S (\Lambda)}{N^{}_S}, \label{eq_eff_mlr} \\
\textrm{FAP}(\Lambda) &\equiv& \frac{N^{}_B (\Lambda)}{N_B}, \label{eq_fap_mlr} 
\end{eqnarray}
where
\begin{eqnarray}
{N}_S(\Lambda) &\equiv& \{x^S_l(\lambda);\lambda \ge \Lambda, l = 1,2,\ldots,N_S\},\\
{N}_B(\Lambda) &\equiv& \{x^B_m(\lambda);\lambda \ge \Lambda, m = 1,2,\ldots,N_B\}, \label{bg_smpls}
\end{eqnarray}
and the ROC curves with Eqs. (\ref{eq_eff_mlr}) and (\ref{eq_fap_mlr}) are drawn in Fig. \ref{roccurve_mlr} by varying $\Lambda$ from $\Lambda=\lambda^B_{\textrm{min}}$ to $\Lambda=\lambda^B_{\textrm{max}}$. From this figure, one can see that the ROC curves with the MLR are more or less similar to the maximum efficiency calculated via the rank given in Fig. \ref{roc_with_rank}. Also, we see that ANNs' performances are improved by 5-10$\%$ compared to the detection statistic as expected from the histogram of Fig. \ref{fig_histogram_mlr}. Therefore, we conclude that the MLR-aided ANN can improve the classification performance.

{
\begin{figure*}[t!]
\centering
	\subfigure[NS-BH model for GRB070714B]{\label{eff_triple_nsbh}\includegraphics[width=0.45\textwidth]{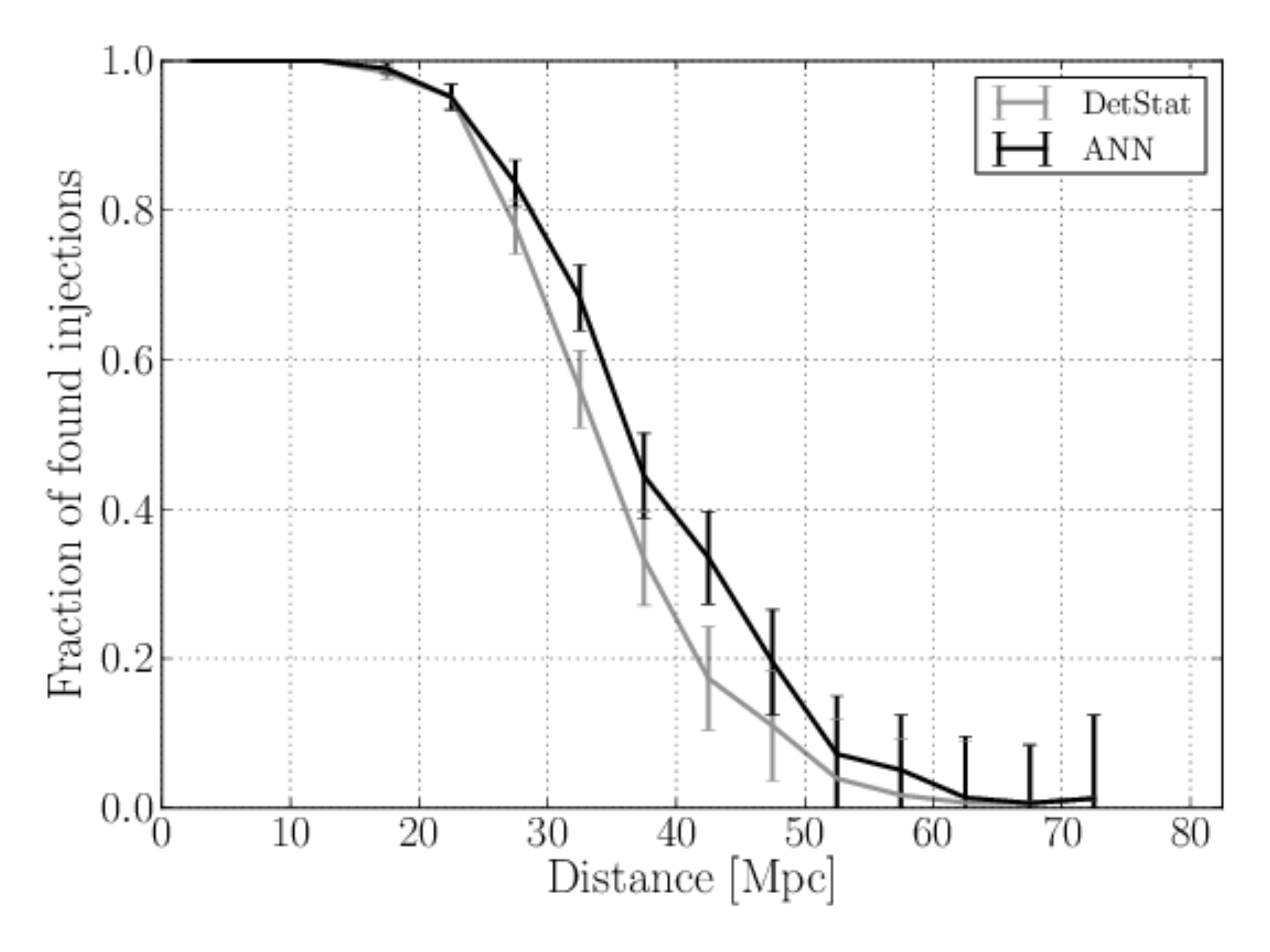}}
	\subfigure[BNS model for GRB070714B]{\label{eff_triple_bns}\includegraphics[width=0.45\textwidth]{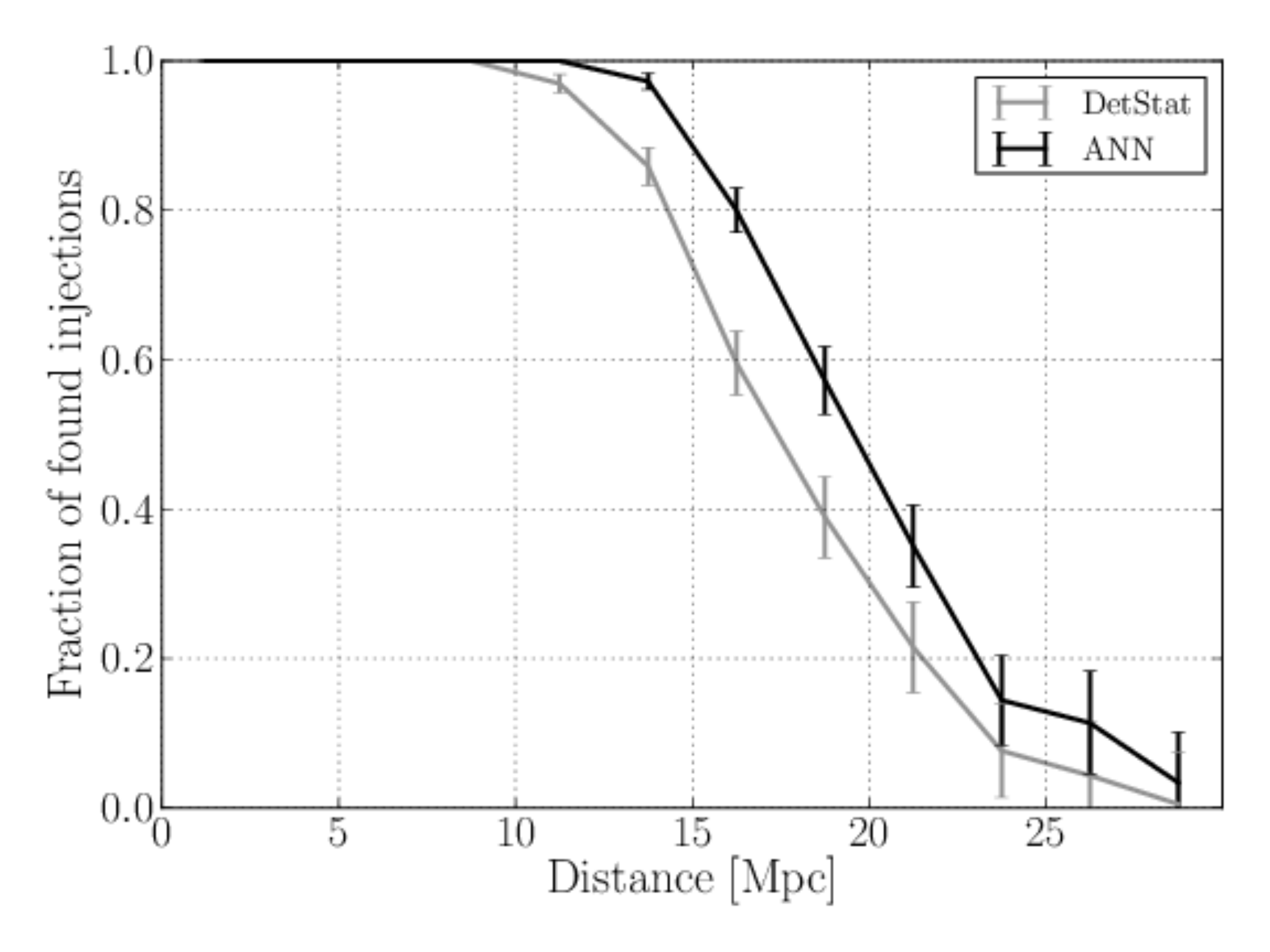}}
	\subfigure[NS-BH model for GRB070923]{\label{eff_triple_nsbh}\includegraphics[width=0.45\textwidth]{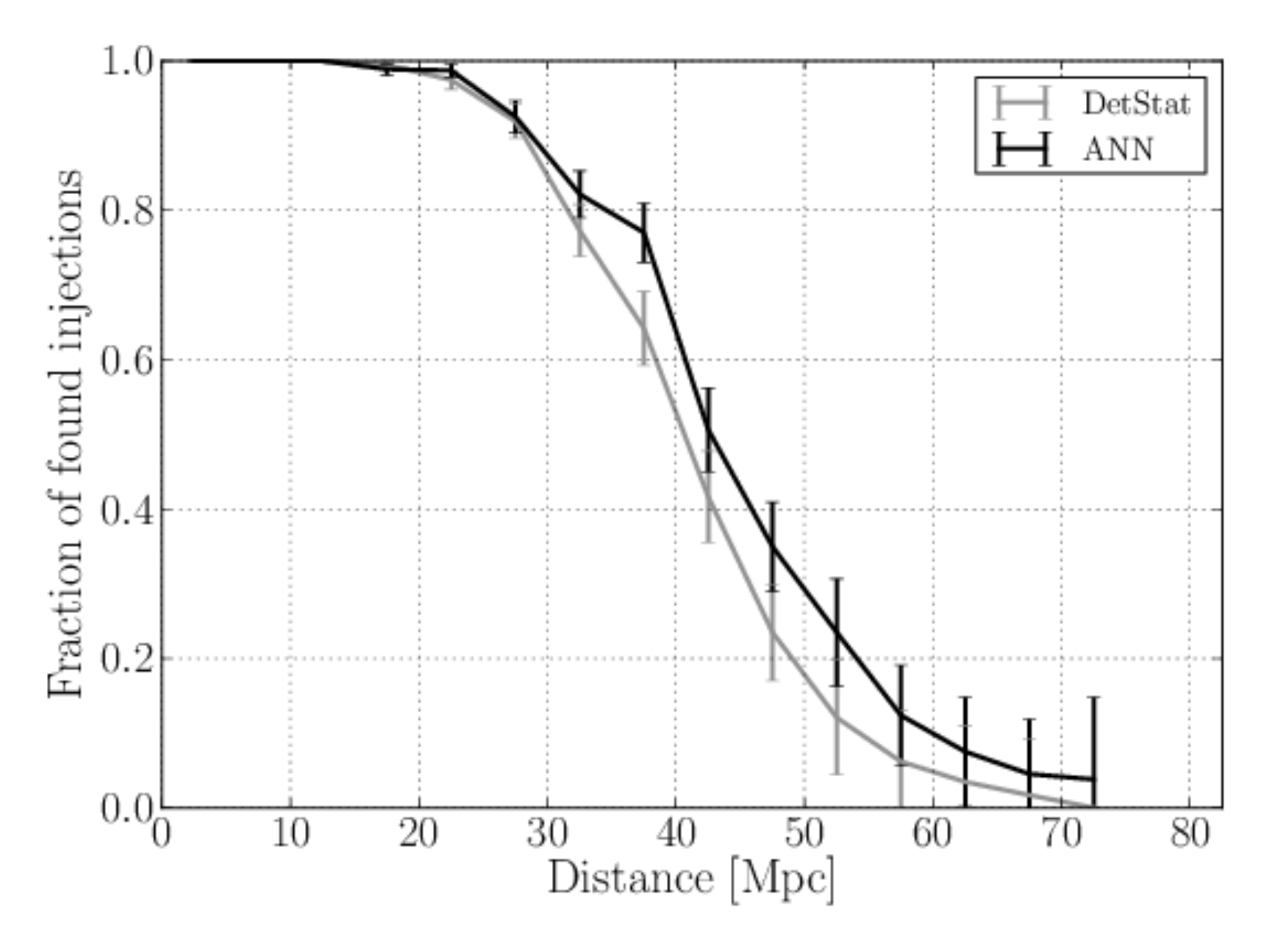}}
	\subfigure[BNS model for GRB070923]{\label{eff_triple_bns}\includegraphics[width=0.45\textwidth]{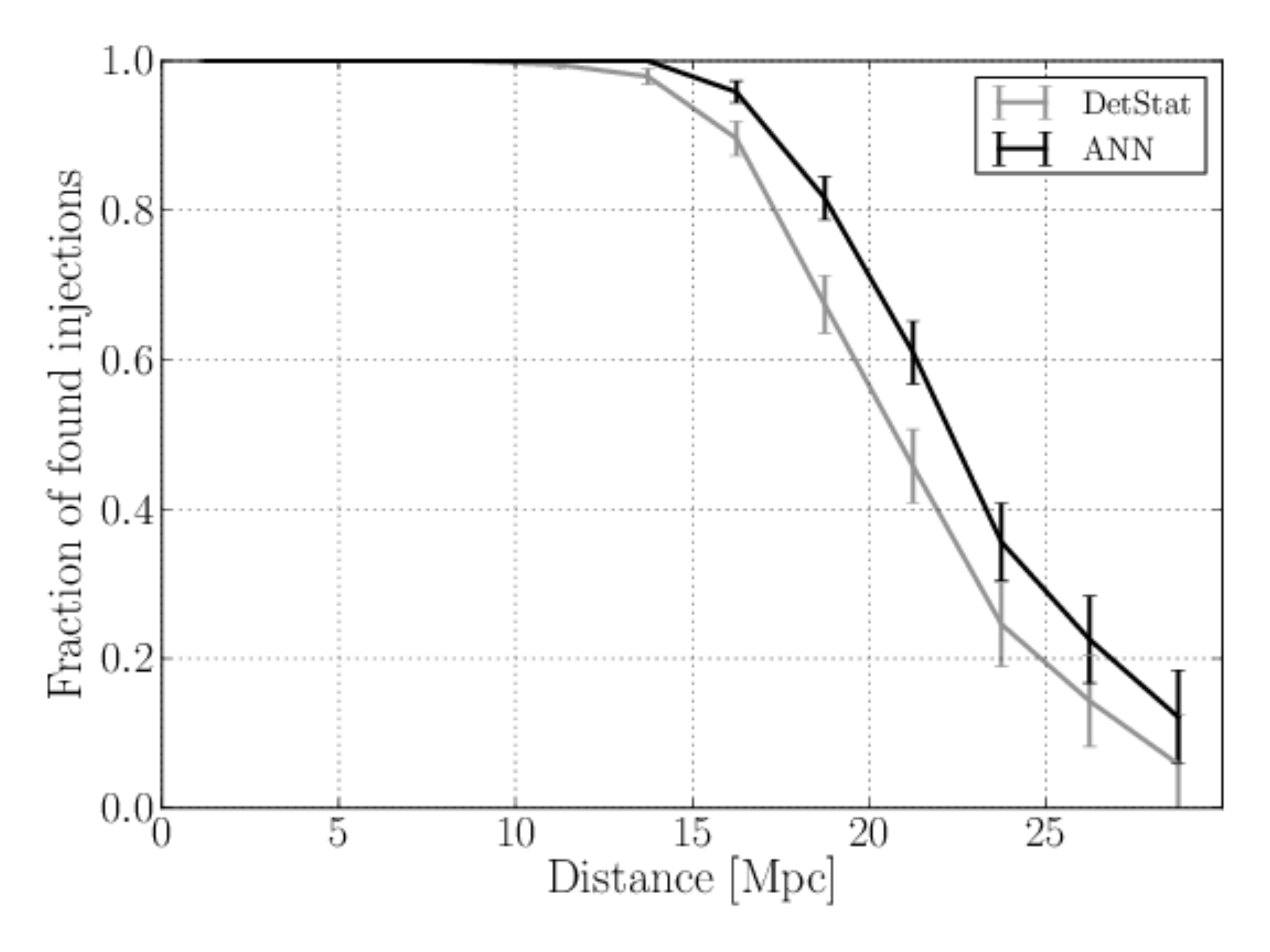}}
\caption{Detection sensitivity of obtaining found injection triggers with respect to the distance to the short GRB's progenitor. The distances used in this plot are the parameters of the software injeciton. We only consider the found injection triggers which have FAPs $\le 0.1\%$. The error bar is drawn with $1$-$\sigma$ confidence level based on the normal distribution.}
\label{fig_detect_eff}
\end{figure*}
}

Meanwhile, when we compare NS-BH model and BNS model of each GRB data, we find that the classification efficiency of the BNS model is better than that of the NS-BH model. To understand this difference, we look the scatter plots of $m_1$ vs. $\chi^2$ in Fig. \ref{plot_features_scatter}. From the scatter plot of NS-BH model, one can see that the $m_1$ parameters of signal and background samples are distributed almost in the same region. On the other hand, the scatter plot of the BNS model shows that the $m_1$ parameters of signal samples have different distribution, that is, rather squeezed distribution compared to the distribution of background samples. From this comparison, we expect that if there are visible differences in the distributions of feature parameters, $m_1$ in this case, the classification performance of ANN is biased by the training samples.

\section{Applications}
\label{applications}

\subsection{Detection Sensitivity on Distance}
\label{DetSense}

So far, we have demonstrated the applicability ANNs to classifying the samples generated by running the CBC-GRB coherent search pipeline on the selected data segments and have demonstrated the ANN's classification performance on the data classification via plotting the ROC curves. 

In this section, we demonstrate the results of performance test in terms of an astrophysical observable, i.e., the distance. It is possible to estimate the detection efficiency as a function of the distance since we have set the distribution of possible distance range for the simulated signals.

Firstly, we define the number of found injection samples that is exceeding a criterion MLR, $\Lambda_T$ in a $l$-th distance bin $[D_l, D_{l+1})$ such as
\begin{eqnarray}
{N_{\textrm{inj}}^{\textrm{found}} (\Lambda_T) |}_{ [D_l, D_{l+1})} = {\{ x^{\textrm{found}}_{\textrm{inj}}(\lambda); \lambda \ge \Lambda_T \} | }_{ [D_l, D_{l+1})}. \label{eq_det_sen_numerator}
\end{eqnarray}
Here, we set $\Lambda_T$ to give $0.1\%$ of FAP (or, equivalently, $3.29$-$\sigma$ confidence level).
With Eq. (\ref{eq_det_sen_numerator}), we define the fraction of correctly classified signal samples, $P$, for each distance bin, $[D_l, D_{l+1})$, such as
\begin{eqnarray}
P(\Lambda_T)|_{ [D_l, D_{l+1})} \equiv \frac{ {N_{\textrm{inj}}^{\textrm{found}} (\Lambda_T) |}_{ [D_l, D_{l+1})} }{ {N_{\textrm{inj}}^{\textrm{tot}}(\Lambda_T) |}_{ [D_l, D_{l+1})} }, \label{detect_eff}
\end{eqnarray}
where the denominator denotes the total number of signal samples in a given distance bin and it is defined as
\begin{eqnarray}
{N_{\textrm{inj}}^{\textrm{tot}} (\Lambda_T)|}_{ [D_l, D_{l+1})} &\equiv& {N_{\textrm{inj}}^{\textrm{found}} (\Lambda_T)| }_{ [D_l, D_{l+1})} + \nonumber\\
&&{N_{\textrm{inj}}^{\textrm{missed}} (\Lambda_T)| }_{ [D_l, D_{l+1})}. \label{N_Stot}
\end{eqnarray}
In order to get the appropriate numbers of both terms, ${N}_{\textrm{inj}}^{\textrm{found}}(\Lambda_T)$ and ${N}_{\textrm{inj}}^{\textrm{missed}}(\Lambda_T)$ in the right-hand-side of Eq. (\ref{N_Stot}), we use the found injection triggers and missed injection triggers, respectively. 
We can see that if there are no missed triggers in a distance bin, namely, ${N}_{\textrm{inj}}^{\textrm{missed}}$ equals zero, then
\begin{eqnarray}
{N}_{\textrm{inj}}^{\textrm{tot}}(\Lambda_T) = {N}_{\textrm{inj}}^{\textrm{found}}(\Lambda_T)
\end{eqnarray}
is easily derived from Eq. (\ref{N_Stot}) and the right-hand-side of Eq. (\ref{detect_eff}) becomes 1. With the calculated $P(\Lambda_T)$, we draw the detection sensitivity vs. distances to the progenitor, as presented in Fig. \ref{fig_detect_eff} for the given distance bins. One can see that the signal samples injected in the close distance range ($\lesssim 10$ Mpc) are mostly found injection events and the number of missed injection events gradually increases as the distance becomes larger. We also draw error bars assuming a binomial distribution with 1-$\sigma$ confidence interval for each plotting point.

{\begin{table}[t!]
\renewcommand{\arraystretch}{1.2}
\caption{Distances where 90$\%$ of injection triggers can be correctly classified by detection statistic and ANNs. As in Fig.\ \ref{fig_detect_eff}, we only consider injection triggers which have FAPs are equal to or smaller than $0.1\%$.}\label{detect_dist_90}
\begin{center}
\begin{tabular}{ l  c  c  c }
	\multirow{2}{*}{Data} & \multirow{2}{*}{Waveform Type} & \multicolumn{2}{c}{Distance at 90$\%$ of Probability} \\
	\cline{3-4}
	{} & {} & \multicolumn{1}{c}{DetStat} & \multicolumn{1}{c}{ANNs} \\
	\hline
	\multirow{2}{*}{GRB070714B} & NS-BH & 23.9 Mpc & 24.5 Mpc \\
	{} & BNS & 12.7 Mpc & 14.8 Mpc \\
	\multirow{2}{*}{GRB070923} & NS-BH & 28.1 Mpc & 28.5 Mpc \\
	{} & BNS & 16.0 Mpc & 17.2 Mpc \\
\end{tabular}
\end{center}
\end{table}}

{
\begin{figure*}[p!]
\centering
	\subfigure[NS-BH model for GRB070714B]{\label{eff_triple_nsbh}\includegraphics[width=0.45\textwidth]{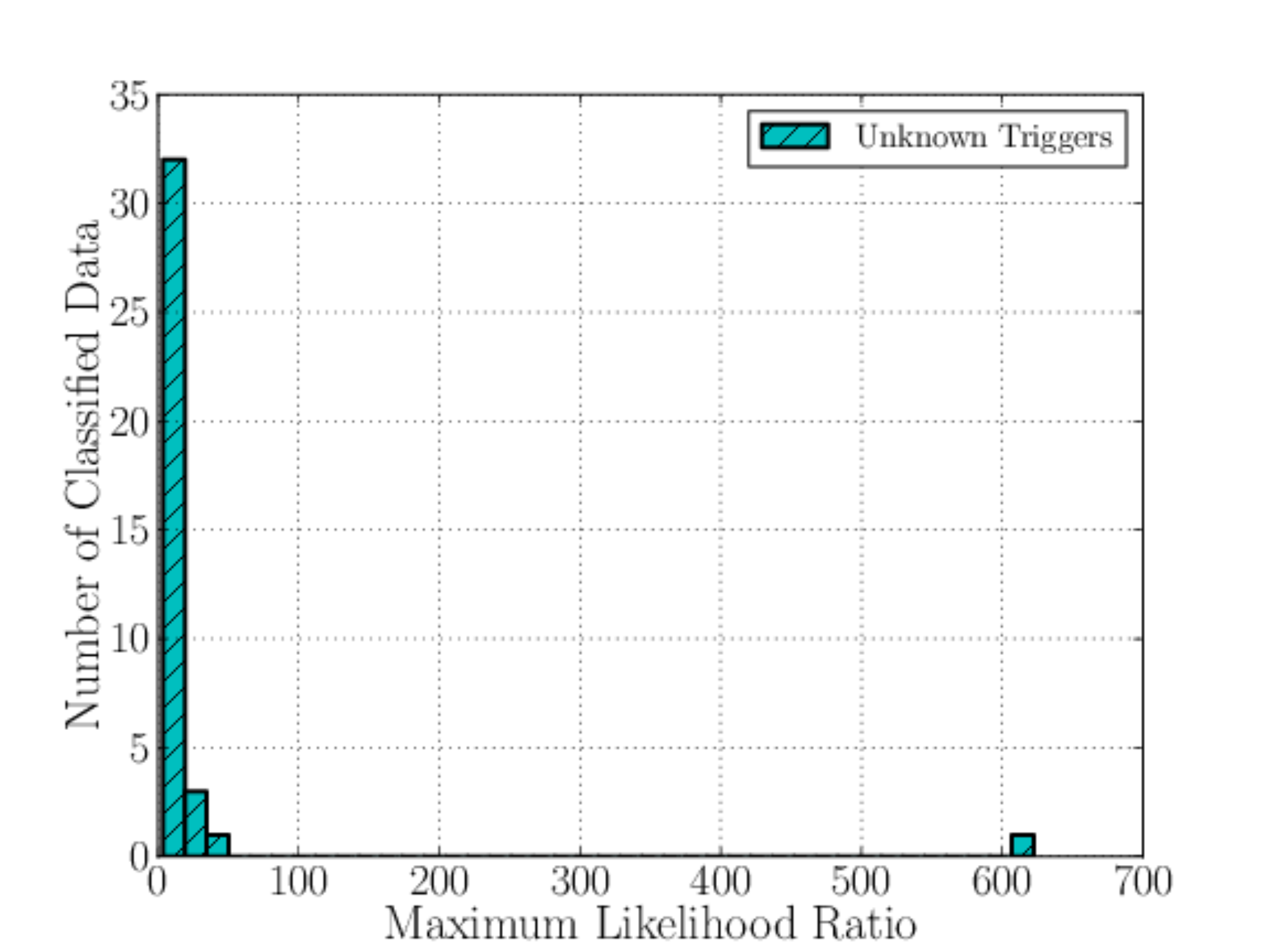}}
	\subfigure[BNS model for GRB070714B]{\label{eff_triple_bns}\includegraphics[width=0.45\textwidth]{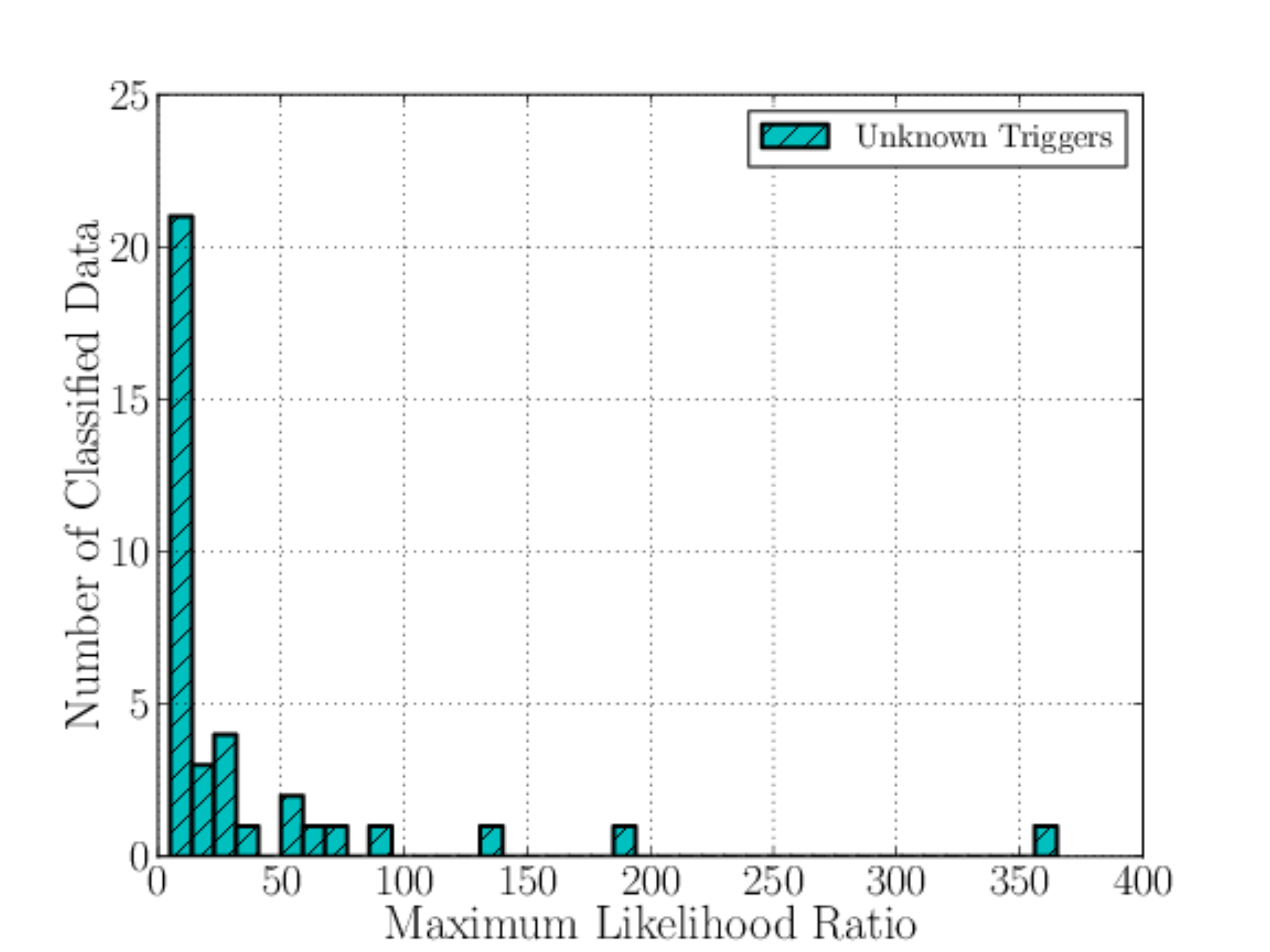}}
	\subfigure[NS-BH model for GRB070923]{\label{eff_triple_nsbh}\includegraphics[width=0.45\textwidth]{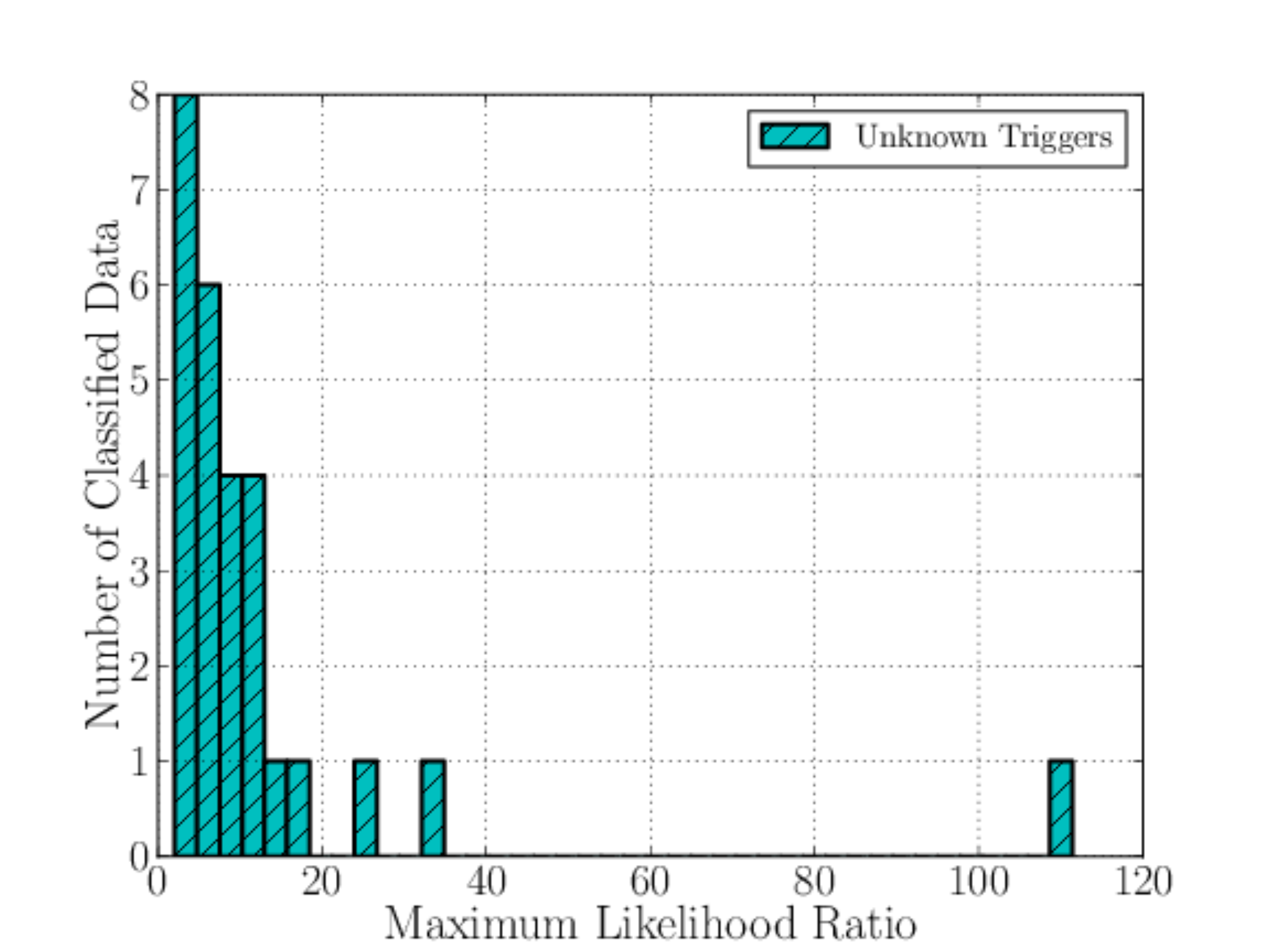}}
	\subfigure[BNS model for GRB070923]{\label{eff_triple_bns}\includegraphics[width=0.45\textwidth]{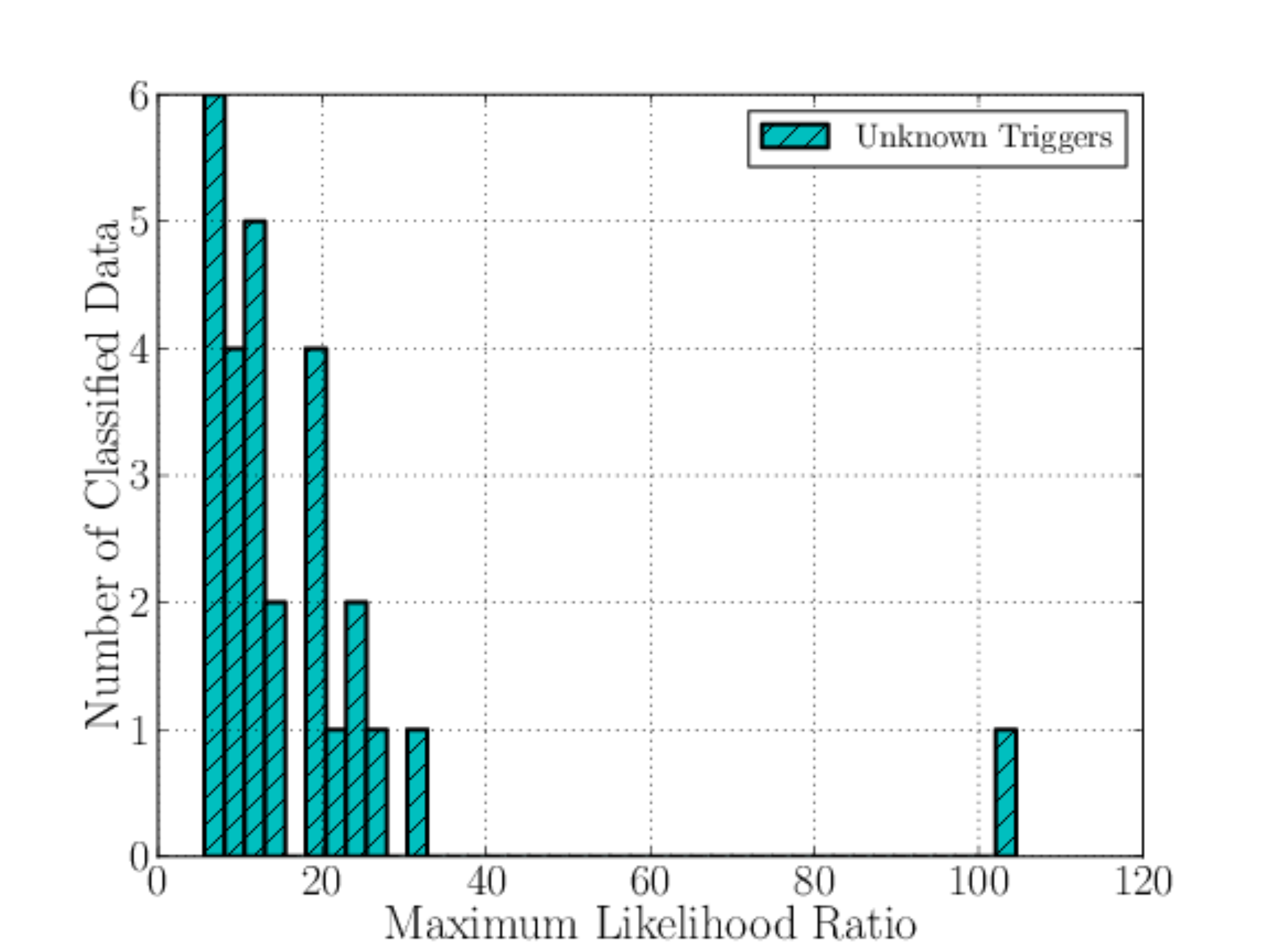}}
\caption{(Color online) The histogram of the maximum likelihood ratios of unknown triggers. For each case, one trigger has a relatively significant value of the MLR than others. The loudest triggers of NS-BH and BNS models for GRB070714B are different such as 6th and 20th triggers among 37 triggers. On the other hand, for GRB070923, the loudest triggers of both NS-BH and BNS models are diffrent too (6th and 14th among 27 triggers).}
\label{fig_histogram_on_trigs}
\end{figure*}
}
{
\begin{figure*}[p!]
\centering
	\subfigure[NS-BH model for GRB070714B]{\label{eff_triple_nsbh}\includegraphics[width=0.45\textwidth]{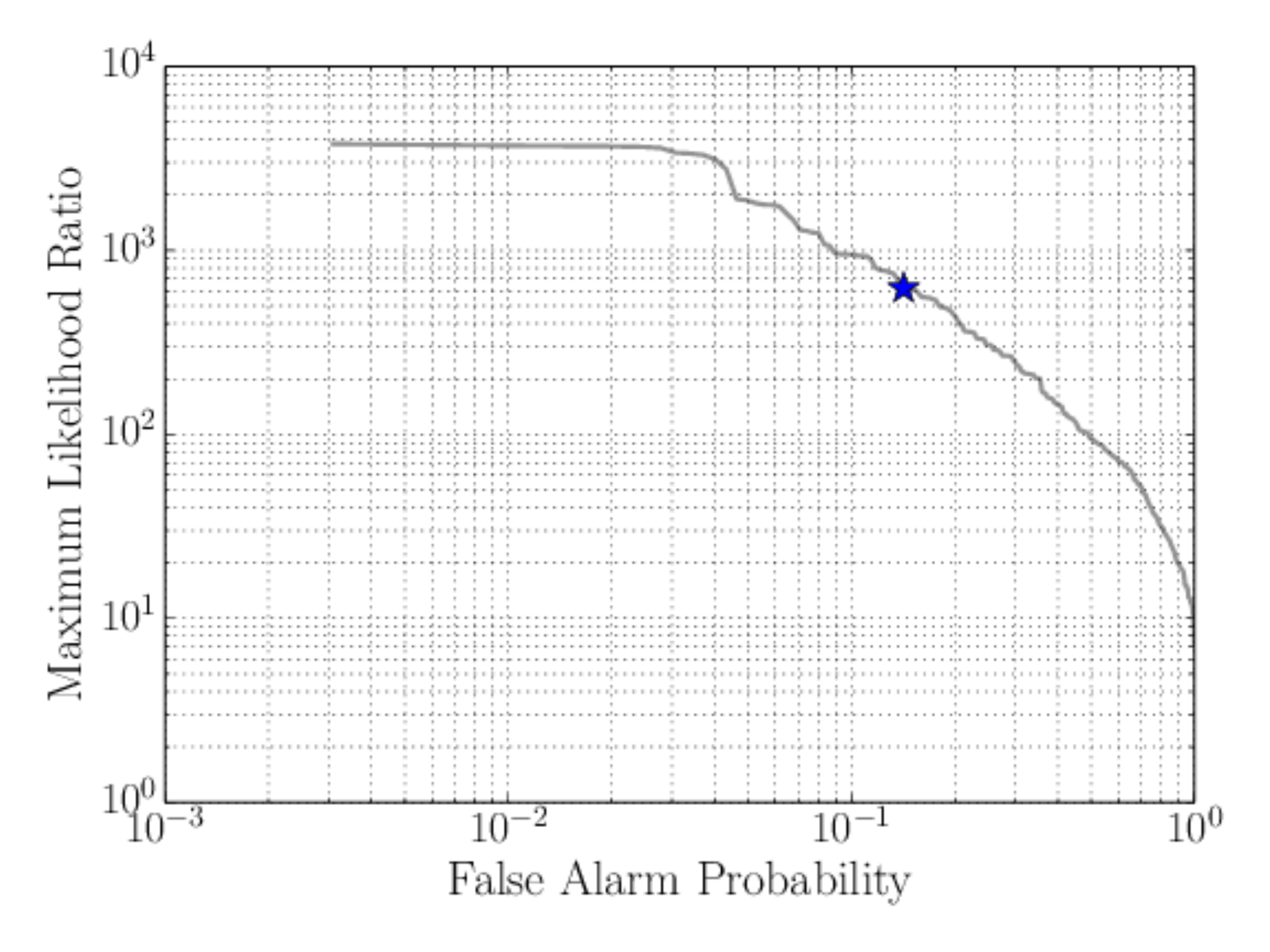}}
	\subfigure[BNS model for GRB070714B]{\label{eff_triple_bns}\includegraphics[width=0.45\textwidth]{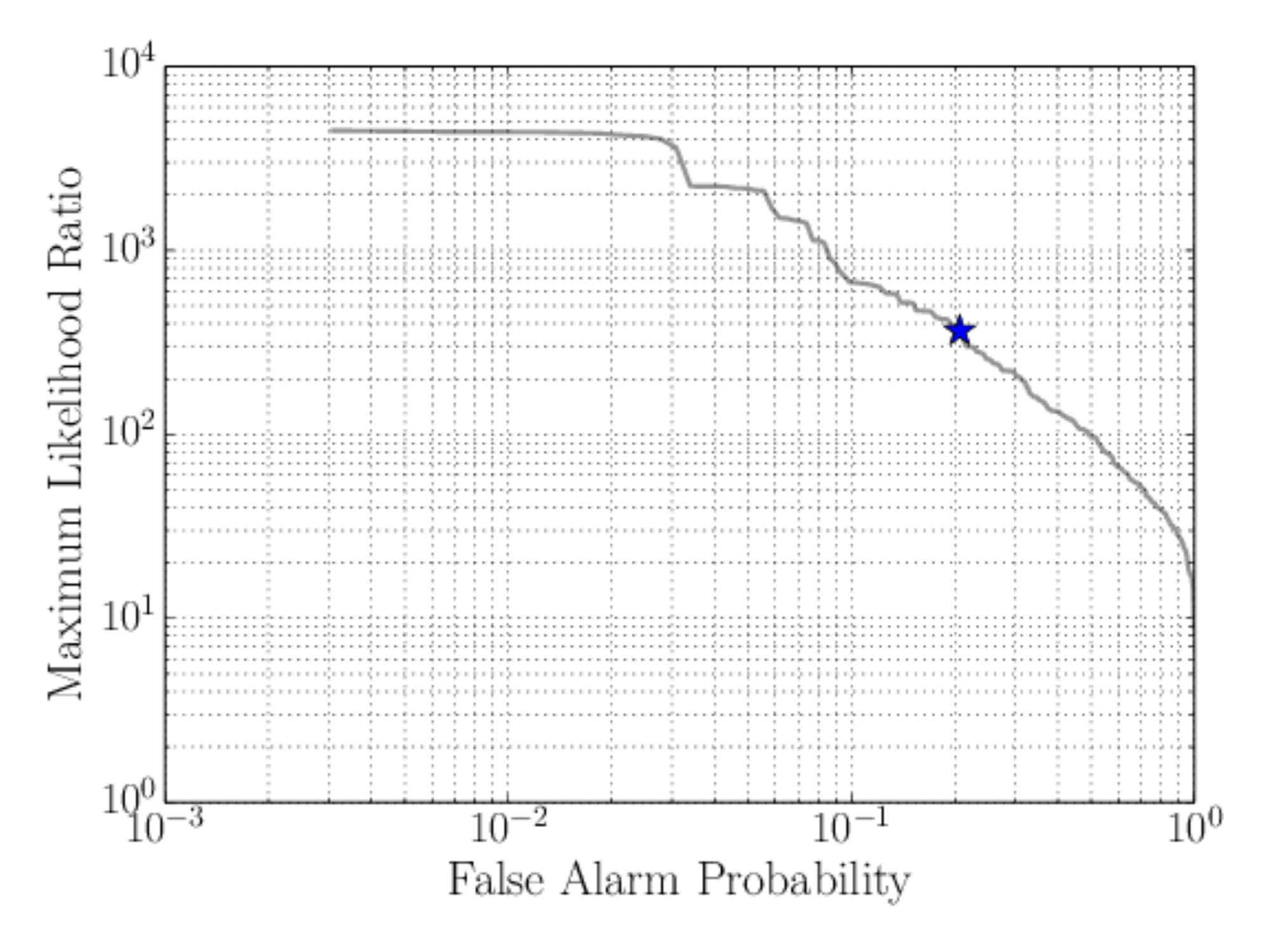}}
	\subfigure[NS-BH model for GRB070923]{\label{eff_triple_nsbh}\includegraphics[width=0.45\textwidth]{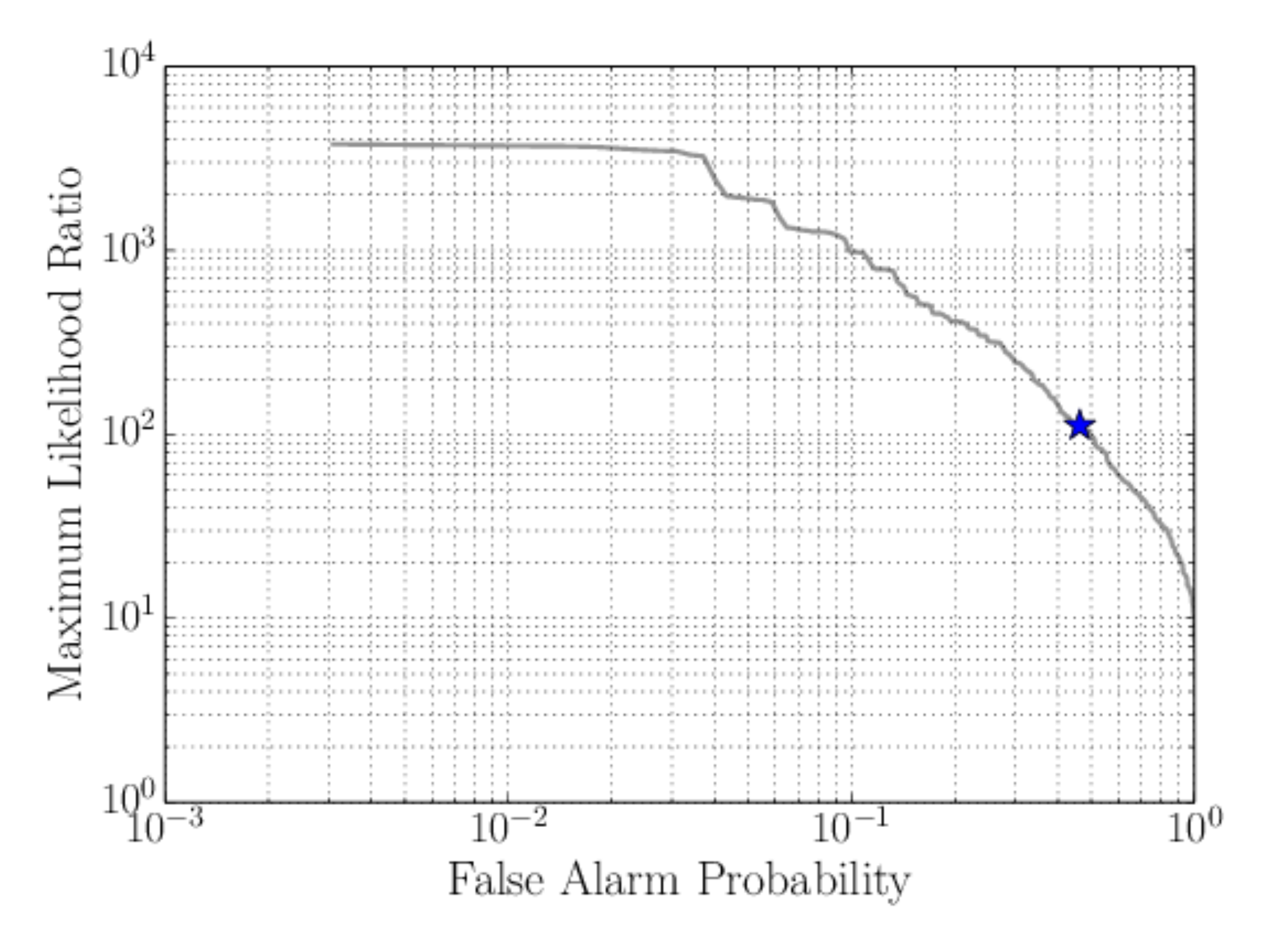}}
	\subfigure[BNS model for GRB070923]{\label{eff_triple_bns}\includegraphics[width=0.45\textwidth]{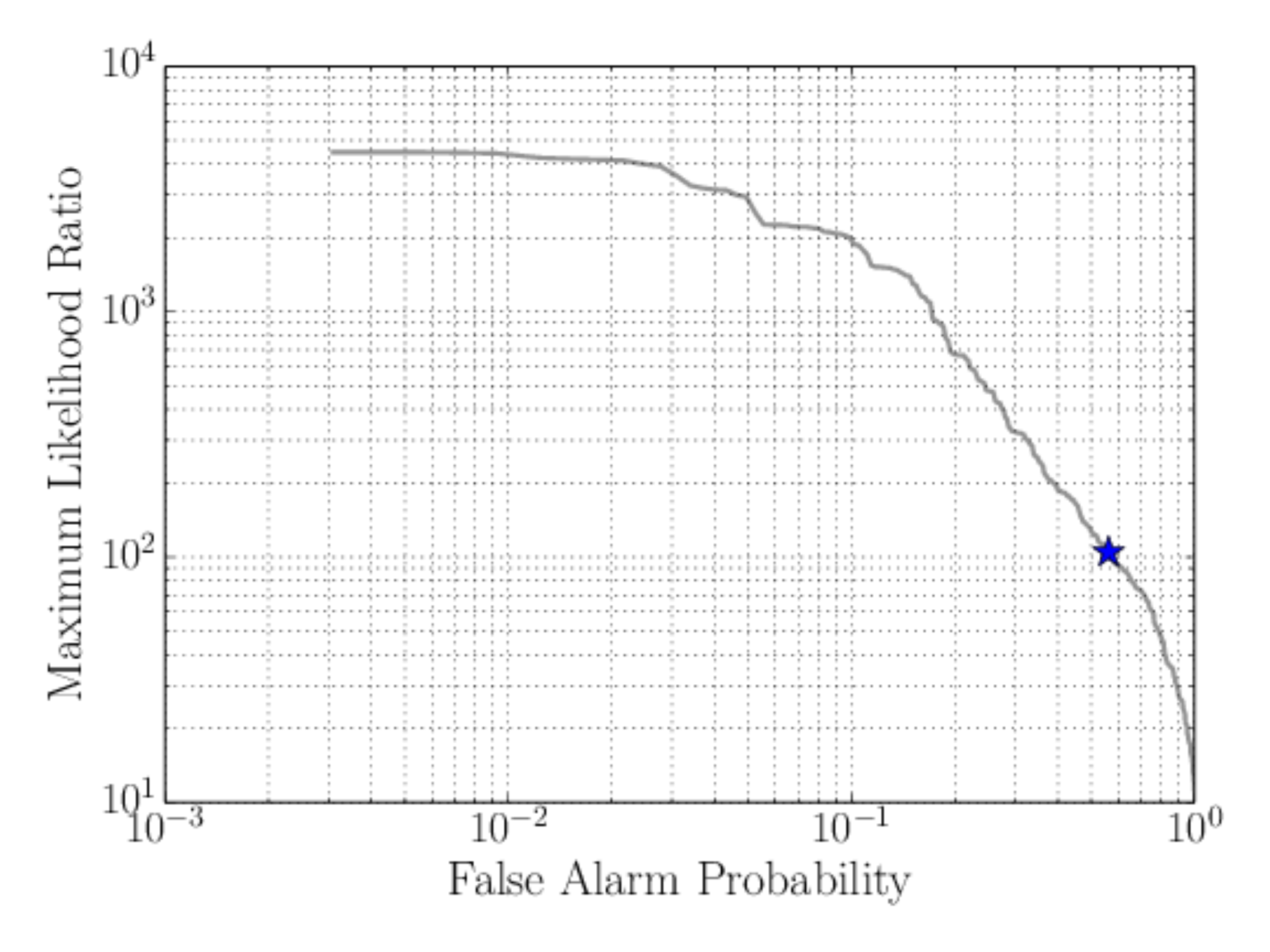}}
\caption{(Color online) The examination of the FAPs of the loudest unknown triggers. The loudest trigger of each case is marked as blue star. The gray lines indicate the FAPs of the background samples. As one can see that the FAPs of the loudest triggers are placed on the line of FAPs of the background samples for both models. This means that those loudest triggers are not GW signals but background events.}
\label{fig_mlr_vs_fap}
\end{figure*}
}

From Fig. \ref{fig_detect_eff}, we clearly see that the ANN is a more efficient tool in detecting short GRBs at large distances for a fixed probability of finding correctly classified signal events. The detectable distances at 90$\%$ of detection efficiency are summarized in Table \ref{detect_dist_90}. The distances in this table mean that, if a short GRB event occurs within this distance, the probability of detecting the GW signal associated with the GRB event is 90$\%$ at least. Thus, if the detection of a GW signal is successful and we can properly recover the distance parameter of the waveform of the GW signal, we can estimate the distance to the progenitor of the observed event, i.e., the common source of the GW signal and short GRB. 

However, the distances summarized in Table \ref{detect_dist_90} are not the exclusion distance to the co-progenitor of a GW signal and a short GRB event. To calculate the exclusion distance, we need to take the errors such as error in the calibration of GW data and error in the amplitude of template waveform into account. Thus, it is hard to directly compare our results to the exclusion distances given in the result of the previous search \cite{Abadie:2010cbcgrb}.

\subsection{Evaluation of Unknown Triggers}
\label{onsource}


We would like to extend this work to the classification of unknown triggers. For this purpose, we use some triggers in a randomly chosen 6 seconds block of the buffer segment\footnote{The buffer segment is adopted to prevent biasing our background estimation due to a potential loud signal in the on-source segment \cite{Fotopoulos:thesis}. Therefore the buffer segment is placed between the on- and off-source segment.} to mimic an on-source segment\footnote{In the real CBC-GRB search, we do not know whether the on-source triggers are real GW signals or noise artifacts. Thus we intend to mock up the situation.} and evaluate them with the trained ANNs in the previous section. We believe that no triggers in the buffer segment are associated with any GRB.

The coherent search pipeline finds 37 triggers and 27 triggers, respectively, in each of the selected blocks in buffer segments of GRB070714B and GRB070923. We evaluate these triggers 100 times with the 100 sets of different connection weights as done in the previous section and then combine their ranks with the MLR. The distribution of calculated MLRs of the unknown triggers are plotted in Fig. \ref{fig_histogram_on_trigs}. 
When we compare this figure with Fig. \ref{fig_histogram_mlr}, it looks like all unknown triggers, even the loudest triggers, can be seen as either signal or background samples for both NS-BH and BNS model cases. From this comparison, we suggest a criterion: if the MLR of the loudest unknown trigger $\lambda_{\textrm{unknown}}$ is smaller than the loudest MLR of background samples $\lambda^B_{\textrm{max}}$, the loudest trigger is close to a background event. On the other hand, if $\lambda_{\textrm{unknown}}$ is greater than $\lambda^B_{\textrm{max}}$, the FAP becomes zero because the numerator of Eq. (\ref{eq_fap_mlr}) becomes zero under this condition.

Meanwhile, the loudest triggers of the NS-BH and BNS models are different (7th and 21st triggers, respectively, among 37 triggers in the GRB070714B buffer segment and 7th and 15th triggers, respectively, among 27 triggers in the GRB070923 buffer segment). We examine these triggers more precisely by estimating the FAPs of the loudest triggers as drawn in Fig. \ref{fig_mlr_vs_fap}. Here, we have changed the entry of background samples from all off-source triggers to the selected off-source triggers which are the triggers having the loudest MLR of each 6 seconds block as done in Ref.\ \cite{Abadie:2010cbcgrb}.\footnote{In Ref.\ \cite{Abadie:2010cbcgrb}, the authors were interested in the existence of GW signal within the 6 seconds on-source segment via examining the most significant trigger. Thus, they divided the off-source segment into 6 seconds long 324 trials to estimate the distribution off background due to the accidental coincidences of noise triggers.}  From this change, the number of reduced background samples, $N'_B$ becomes 324 and Eq. (\ref{bg_smpls}) changes to 
\begin{equation}
{N}'_B(\Lambda') = \{x^B_q(\lambda');\lambda' \ge \Lambda', q = 1,2,\ldots,324\}.
\end{equation}
In Fig. \ref{fig_mlr_vs_fap}, we draw the FAPs of background samples as gray lines using Eq. (\ref{eq_fap_mlr}) by varying from ${\Lambda}'=\lambda'^{B}_{\textrm{min}}$ to $\Lambda'=\lambda'^{B}_{\textrm{max}}$. From these figures, we see that the FAPs of the loudest unknown triggers are placed on the line of FAPs of the background samples. If they were real GW signals, the loudest triggers should be placed out of the line of FAPs of the background samples, i.e., it should be placed in more left and upper area of the minimum FAP.\footnote{For this case, we need more precise follow-up analysis, e.g., testing correlation with known background events.} This result shows that those loudest unknown triggers are less significant than the loudest background sample and are not GW signals. Also, the calculated FAPs are $0.14$ (NS-BH) and $0.21$ (BNS) for the loudest triggers in the GRB070714B buffer segment and $0.46$ (NS-BH) and $0.56$ (BNS) for the loudest triggers in the GRB070923 buffer segment. Therefore, we conclude that we find no significant trigger in the selected 6 seconds block.

\section{Summary and Discussion}
\label{summary}

In this work, we discuss the improvement of the search performance for GW candidate events related to short GRBs by using ANN algorithm compared to a conventional detection statistic. With this demonstration, we aim to increase the search sensitivity on GWs associated with short GRBs. 

We use the GW data obtained by the LIGO and Virgo detectors during S5 and VSR1 and take short GRBs, GRB070714B and GRB070923 as test samples. By using the coherent CBC-GRB search pipeline, on-source, off-source, and software injection triggers are generated. For the generation of the software injection triggers, we consider both NS-BH and BNS binaries as the progenitor of a GRB for the determination of component masses. We set the distributions of the distances to the progenitors to have different ranges depending on the detectors' responses at the given event time and/or on the type of binary system. We train ANNs with taking found injection triggers as signal samples and off-source triggers as background samples. Then, we evaluate test samples with the trained ANNs. Each sample for both training and evaluation is characterized by the 8 statistical quantities and 2 physical quantities listed in Table \ref{features}.

The training process is done by minimizing the error between the observed value from a given ANN and the target values of samples. In this work, the error is calculated by the mean-squared-error (MSE), which is defined in Eq. (\ref{eq_mse}). For this result, one can suspect that the training processes were stuck in local minima. Recent work \cite{Pascanu,Dauphin} show that there is no such distinct local minima compared to the global minimum in the higher dimensional non-convex optimization.

It appears however that the performance on data classification vary significantly between the trained ANNs. We find that this variation results from the statistical variance in the ANN algorithm because ANN uses a randomly distributed initial input configuration. Therefore, we need to mitigate the variation in order to get a reliable interpretation on our results. For the mitigation of the statistical variance, there are several possible ways of reducing the variance, for example, calculating ensemble average of ranks, taking median value, and computing maximum likelihood ratio (MLR). Among those, we adopt MLR because the other two methods are not suitable to the data as shown in Fig. \ref{rank_hist_of_a_sample}. Moreover, the MLR method has already been shown to be an optimal method in obtaining a representative quantity for GW data in Refs. \cite{Biswas:2013prd} and \cite{Biswas:2012prd}. With this prescription, we resolve the variance in ranks and the classification performance by combining the results from 100 trials. 

As a result of the performance test, we see that the background samples have finite values of MLR from the distribution of MLRs given in Fig.\ \ref{fig_histogram_mlr}. However, for signal samples, about half of samples have similar values of MLR to the values of background samples and the rest of them are separated from the range of finite MLRs and take an infinite value. This tendency is consistently shown in all considered cases. When we compare this Fig. \ref{fig_histogram_mlr} to the histogram of ranks in Fig. \ref{rank_hist_of_all_samples}, we see that the classification efficiency with the MLR is enhanced by the fraction of clearly separated signal samples. 

Also, we examine the improvement in the classification performance by comparing the ROC curves, instead of directly comparing the ranks of two ranking methods, because there is no rule to connect two different ranks, scored independently by the conventional detection statistic or by the ANN. When we look the ROC curves, we find that the ROC curve obtained by MLR are more or less similar to the maximum efficiency calculated via the ANN's rank given in Fig. \ref{roc_with_rank}. From the comparison of the ROC curves between the one obtained by using the conventional detection statistic and the other one obtained by using the ANN, we see that the data classification efficiency at the minimum false alarm probability (FAP) is improved by 5\% -- 10$\%$ as expected from the histogram of Fig. \ref{fig_histogram_mlr}. Therefore, we conclude that the MLR-aided ANN can improve the classification performance.


Meanwhile, when we compare NS-BH model and BNS model of each GRB data, we find that the classification efficiency of the BNS model is better than that of the NS-BH model. To understand this difference, we look the scatter plots of $m_1$ vs. $\chi^2$ in Fig. \ref{plot_features_scatter}. From the scatter plot of NS-BH model, one can see that the $m_1$ parameters of signal and background samples are distributed almost in the same region. On the other hand, the scatter plot of the BNS model shows that the $m_1$ parameters of signal samples have different distribution, that is, rather squeezed distribution compared to the distribution of background samples. From this comparison, we expect that if there are visible differences in the distributions of feature parameters, $m_1$ in this case, the classification performance of ANN is biased by the training samples. 

For the real search, we need to support the identification of a GW signal related to a short GRB event by recovering thel distance to the progenitor of GWs. With the help of the observations on GRBs during recent decades, one can estimate the redshift of the event and estimate the distance to the source. However, the redshift of short GRBs is not determined well because the duration time of afterglow of a short GRB is not long enough to localize the position of the host galaxy which is need to perform spectroscopic analysis for measuring relevant redshift \cite{Rowlinson:2010mnras}. Also, the uncertainty on the sky location occurred by the large field of view of the GRB telescope such as \emph{Fermi} satellite \cite{Meegan:2009} hinders the accurate measurement of the short GRBs' redshifts. So, we know the redshift values only for several short GRB events, e.g., 4 GRBs out of 22 GRBs analyzed for the Ref. \cite{Abadie:2010cbcgrb}. Thus, if we can detect a GW signal related to a short GRB and if the detection efficiency as a function of the distances can be provided, we may obtain more accurate information on the distance to the source in addition to the redshift measurement. 

We extend the results of the MLR-aided ANN's classification performance to the estimation of the sensitivity of analysis as a function of the distance which denote the upper bound of the distance where we can observe a GW event within the given distance. In this work, it is possible to estimate this quantity since we have set the distribution of possible distance range for the simulated signals. For this estimation, we assume that mass parameters are marginalized. From the results presented in Fig. \ref{fig_detect_eff} and Table \ref{detect_dist_90}, we find that the sensitivity of analysis as a function of the distance at 90\% of probability can be increased by $\sim$ 1.4\% -- 16.5\% compared to the distance calculated with the conventional detection statistic. This means that if a GW event occur at a distance which are greater than the distance estimated by the conventional detection statistic and smaller than the distance estimated by the MLR-aided ANN, we can identify that event with MLR-aided ANN. Therefore, it is shown that the estimated sensitivity of analysis obtained by using the MLR-aided ANN's classification results allows us to observe more events occurring at farther distance than the conventional method. However, as discussed in Section \ref{DetSense}, the distances summarized in Table \ref{detect_dist_90} are not the exclusion distance to the co-progenitor of a GW signal and a short GRB event. To calculate the exclusion distance, we need to take the errors such as error in the calibration of GW data and error in the amplitude of template waveform into account. Thus, it is hard to directly compare our results to the exclusion distances given in the result of the previous search \cite{Abadie:2010cbcgrb}.


We apply our analysis to the data in the buffer segment of the two GRBs. From the evaluation of triggers in the buffer segment, we see that the loudest triggers are placed on the line of FAPs of the background samples as shown in Fig. \ref{fig_mlr_vs_fap}. If they were real GW signals, the loudest triggers should be placed out of the line of FAPs of the background samples, i.e., it should be placed in more left and upper area of the minimum FAP. Therefore, this result shows that those loudest triggers are less significant than the loudest background sample and they cannot be GW signals. The estimated FAPs are $0.14$ (NS-BH) and $0.21$ (BNS) for the loudest triggers in the GRB070714B buffer segment and $0.46$ (NS-BH) and $0.56$ (BNS) for the loudest triggers in the GRB070923 buffer segment. Therefore, it can be concluded that we find no significant event with the analysis with MLR-aided ANN on the buffer triggers.

In this analysis, the lowest obtainable FAP, that is defined by one over the number of background samples which are determined by the pipeline, is found to be much higher ($\sim 10^{-3}_{}$) than the required value ($\le 10^{-7}$) for a clear declaration of a GW signal. In order to lower the minimum FAP to the expected level, we need to have at least 10,000 times more background samples with such a method given in \ref{fap_lowering}: adding more segments not far from the original segment by seeking additionally available segments via shifting original GRB's event time by sidereal days to get segments with the fixed sky location because we may assume that the profile of background transients around given sky location are not much differ from the originally interesting segment. As an alternative method of getting additional background samples, (i) time slides of GW data \cite{S5LowMassLV} or (ii) extrapolating given distribution of FAP such as done in Ref. \cite{Adams:2013prd} can be adopted. It remains as a future work how to implement those methods in the analysis.

Throughout this work, we investigate the feasibility of application of ANN to CBC-GRB search as a new ranking method and we find that it can improve the search performance by estimating the FAP and the detectable distance. Therefore, we would suggest that the artificial neural network can be a complementary method to the conventional detection statistic for identifying gravitational-wave signals related to the short gamma-ray bursts.

We however have a limit obviously in arguing the robustness and the consistency of this approach since we have tested only four-cases (two GW data $\times$ two binary models). This fact means that we need to test more data. As a possible way of obtaining more test data, we may consider to use GW data for other targeted GRBs summarized in Ref.\ \cite{Abadie:2010cbcgrb,Abadie:2012grb,Aasi:2014IPNgrb}. Or, alternatively, we may choose random sky locations and random times to generate triggers, instead of restricting our focus to known GRBs' event times and sky locations, because we know that GRBs are isotropically distributed in the sky. So, in a future work, we will test more data to examine the robustness of this approach and will discuss the general characteristic of this analysis method.

{\begin{table}[t!]
\renewcommand{\arraystretch}{1.3}
\caption{The horizon distance and volume-weighted average distance range of each detector. All tabulated values are given in unit of Mpc. For the horizon distances of H1 and L1, we take the maximum value of given range in Ref. \cite{LIGOHorizonDistance}. Meanwhile, we read the value of mode from Fig. 1 of Ref. \cite{S5VSR1Sensitivity} for the horizon distance of V1. Among the values of volume-weighted average distance ranges, asterisks indicate the decisive distances.}\label{distances}
\begin{center}
\begin{tabular}{ l c c c c c c}
	\multirow{2}{*}{Data} & \multicolumn{3}{c}{Horizon Distance, $D_{h}^{\textrm{IFO}}$} & \multicolumn{3}{c}{Volume-Weighted Average Distance, $\tilde{D}^{\textrm{IFO}}$} \\
	\cline{2-4}
	\cline{5-7}
	{} & H1 & L1 & V1 & H1 & L1 & V1\\
	\hline
	GRB070714B & 35.0 & 35.0 & 7.5 & $9.1^*$ & 12.9 & 6.2 \\
	GRB070923 & 35.0 & 35.0 & 7.5 & $11.1^*$ & 13.7 & 5.2 \\
\end{tabular}
\end{center}
\end{table}}

\ack
We thank the LIGO Scientific Collaboration and the Virgo Collaboration for the use of the data. We are also grateful for computational resources provided by the Leonard E Parker Center for Gravitation, Cosmology and Astrophysics at University of Wisconsin-Milwaukee (NSF-0923409). The authors would like to thank S. Bose, K. Cannon, T. Dent, P. Graff, C. Hanna, H. M. Lee, C. Kim, and R. Vaulin for helpful comments and useful discussions. KK would like to specially thank J. Burguet-Castell, A. Dietz, and N. Fotopoulos for suggesting the initial motivation of this work. KK, YMK, CHL, HKL, JJO, SHO, and EJS were also supported in part by the Global Research Network program of the National Research Foundation funded by the Ministry of Science, ICT, and Future Planning of Korea (NRF-2011-220-C00029). 

\appendix

\section{Physical Parameters of Simulated Waveforms for Software Injection}
\label{app_SoftInjParams}

In this section, we need to corroborate some physical parameters such as mass, distance, and spin that are used in the generation of simulated waveforms. These parameters represent various types of GW progenitors and the property of waveform itself. For the mass parameter, we consider two appropriate binary systems such as BNS and NS-BH as the sources of simulated signals based on the predicted event rates \cite{ratesdoc}. For the neutron stars in the BNS system, their masses are given in range of $1$-$3$ $M_{\odot}$ with mean mass, ${\overline{m}}_{\textrm{NS}}$ = 1.4 $M_{\odot}$ and standard deviation, $\sigma_{\textrm{NS}}$ = 0.2 $M_{\odot}$ with the assumption of the Gaussian distribution. Similarly, for the NS-BH system, the ranges of component masses are set as $m_{\textrm{NS}}=1$-$3$ $M_{\odot}$ (${\overline{m}}_{\textrm{NS}}$ = 1.4 $M_{\odot}$ for neutron star and $\sigma_{\textrm{NS}}$ = 0.4 $M_{\odot}$) and $m_\textrm{BH}$ 2-25 $M_{\odot}$ (with ${\overline{m}}_{\textrm{BH}}$ = 10 $M_{\odot}$ and $\sigma_{\textrm{BH}}$ = 6.0 $M_{\odot}$) for black hole under the assumption of the Gaussian distribution too (for details, see Ref. \cite{Abadie:2012grb}). 

One can expect from the name of used waveforms for this software injection that they contain spin effect on the contrary to the template waveforms for the matched filtering process. With observed information and models \cite{Mandel:2009nx,Hessels:2006} we assume possible ranges of spin magnitudes for NS and BH as [0,0.4] and [0,0.98), respectively. Therefore, we set possible values of them to be uniform in given ranges with random orientation. 

When we take the spinning waveform into account, the related important parameter for describing a binary system is the inclination angle which describes size of angle between the direction of the total angluar momentum and the line of sight to the observer because the strength of a GW to a detector can be varied depending on the angle. Many observations and models suggest that a GRB is generated by cone-shaped outflow from a CBC system \cite{2006ApJ...653..468B,2006ApJ...653..462G,Dietz:2010eh}. From these, it is supposed that when the inclination angle is placed within the cone it is possible to find a GW signal directly related to a short GRB. Thus, the pipeline uses four different half-opening angles, $10^\circ, 30^\circ, 45^\circ$, and $90^\circ$ as possible sizes of cone and it allows the inclination angle to be distributed within the cone (for more details refer to Ref. \cite{Abadie:2012grb}). We only take $10^\circ$ for simplicity.

The range of distances to the progenitors are differently determined by the type of progenitor and applied with taking the detectors' sensitivities on given GRB's sky location into account too. To estimate the sensitivity, first, we need to calculate the antenna response, ${\cal{F}}$ \cite{Allen:2005fk}:
\begin{equation}
{\cal{F}} = \left( F_+^2 + F_\times^2 \right)^{1/2}, \label{antenna}
\end{equation}
where $F_+$ and $F_\times$ are the antenna factors of a detector for the `$+$'- and `$\times$'-polarizations of a GW signal, respectively (for more details see Ref. \cite{Allen:2005fk}). These $F_{+,\times}$ represent the amount of sensitivity on each polarization of the incident GW signal. 
A GRB's event time and its sky location play an important role in the determination of the values of $F_{+,\times}$. The determined values of $F_{+,\times}$ and ${\cal{F}}$ of each detector for the selected GRB are summarized in Table \ref{GRBs_IFO}. In particular, if the value of ${\cal{F}}$ is 1, one finds that the detector is at mostly optimal location for the putative GW source at that location. On the other hand, 0 value means that it is impossible to see any GW signals with that detector.
When we obtain the antenna response ${\cal{F}}$ of each detector, we can calculate the volume-weighted average distance range $\widetilde{D}^{\textrm{IFO}}$ \cite{FinnChernoff:1993} by simply multiplying ${\cal F}$ to the horizon distance $D_h^{\textrm{IFO}}$ \cite{LIGOHorizonDistance,S5VSR1Sensitivity}
\begin{equation}
\widetilde{D}^{\textrm{IFO}} = {\cal{F}}^{\textrm{IFO}} \times D_h^{\textrm{IFO}}, \label{reweightedD}
\end{equation}
where IFO of the superscripts stands for interferometric observatories, H1, L1, or V1. The horizon distances $D_h^{\textrm{IFO}}$ and volume-weighted average distances $\widetilde{D}^{\textrm{IFO}}$ of relevant detectors for GRB070714B and GRB070923 are listed in Table \ref{distances}. 
From the values of $\widetilde{D}^{\textrm{IFO}}$, we can read the upper limit of reachable distance of each detector. That is, if a GW event is occurred within the distance limit at the given sky location, we may detect that event.  Among the listed $\widetilde{D}^{\textrm{IFO}}$ in Table \ref{distances}, in particular, we choose the second largest value as the decisive distance for the GRB070714B as done in Ref. \cite{Abadie:2012grb}.
Resultantly, applied distance ranges for the software injection are 2-30 Mpc for BNS system and 2-72.5 Mpc for NS-BH case. We also suppose that the sources of GWs are distributed in uniform within the ranges.

\section{Lowering the Minimum False Alarm Probability}
\label{fap_lowering}

For the issue of the lowering the minimum FAP, we may consider a method suggested by Ghosh {\it et al.} \cite{Ghosh:2013} to increase the number of background samples. The main feature of suggested method is adding more segments not far from the original segment. To take account this, they assume that the profile of background transients around given sky location are not much differ from the originally interesting segment. With this method, we seek additionally available segments via shifting original GRB's event time by sidereal days to get segments with the fixed sky location. From the shifting 10 sidereal days, such as from 1- to 5-days before and after the original event time of GRB070714B, we find that there are 6 more available segments. For the availability of segments, we check whether each detector includes the extended data or not. The information about additional segments are summarized in Table \ref{additional_seg}. From this table, one can see that the availability of additional segments are tested based on the type of GW detectors' network on each of the shifted sidereal days. As a result, we obtain at least 6 times more background samples and find that it is possible to lower the minimum FAP to $\sim$$10^{-5}$ ($\sim$$4.4$-$\sigma$) with the increased number of background samples.
{\begin{table}[t!]
\renewcommand{\arraystretch}{1.2}
\caption{List of available segments for GRB070714B. The day shifts are done based on the sidereal day. Because the GW data for this GRB was a triple coincident data of H1, L1, and V1, the available additional data segments are only 6 segments of the day-5, -3, -1, +1, +2, and +5.}\label{additional_seg}
\begin{center}
\begin{tabular}{ l  c  c  }
	Day & UTC Time & Detected IFOs \\
	\hline
	-5 & 2007-07-09 T05:19:08.54180 & H1, L1, V1 \\
	-4 & 2007-07-10 T05:15:12.63344 & H1, L1 \\
	-3 & 2007-07-11 T05:11:16.72508 & H1, L1, V1 \\
	-2 & 2007-07-12 T05:07:20.81672 & H1, V1 \\
	-1 & 2007-07-13 T05:03:24.90836 & H1, L1, V1 \\
	+1 & 2007-07-15 T04:55:33.09164 & H1, L1, V1 \\
	+2 & 2007-07-16 T04:51:37.18328 & H1, L1, V1 \\
	+3 & 2007-07-17 T04:47:41.27492 & H1, V1 \\
	+4 & 2007-07-18 T04:43:45.36656 & H1, L1 \\
	+5 & 2007-07-19 T04:39:49.45820 & H1, L1, V1 \\
\end{tabular}
\end{center}
\end{table}}

On the other hand, Clark \emph{et al.} \cite{Clark:2014} also suggest that time slides of GW data could be used as an alternative method of getting additional background samples.

\section{Round-Robin Process}
\label{round-robin}



In the preparation of input samples, we conduct a pre-process, called \emph{round-robin process}, on the sample data before performing training process. The purpose of this pre-process is to mitigate the possibility of overestimation (or, equivalently, overtraining) which may be occurred by inadequate training with small or limited numbers of sample data. In order to successfully implement this pre-process and reduce the rate of overtraining, we prepare $M$ sets of round-robined sample data by evenly dividing the whole samples of $X_S$ and $X_B$ into $M$ different sets of $X_S^k$s and $X_B^k$s, where $k=1,2,\ldots,M$, in the same manner: a set to be consistently composed of one-tenth of total signal samples and one-tenth of total background samples. When it is done, there are no overlaps between the samples in one set and the samples in other sets. Then, to make a pair of training and evaluation sets, we let one of total round-robined sets to be an evaluation set and rest of them be the pairing training set. Then, with the definition in Eqs. (\ref{eq_sig_samples}) and (\ref{eq_bg_samples}), we can get the first pair of an evaluation set $E^{1\textrm{st}}_{}$ and a training set $T^{1\textrm{st}}_{}$ as follows:

\begin{eqnarray}
E^{1\textrm{st}}_{} &=& X^{1}_S \cup X^{1}_B \nonumber \\
                          &=& \{x_S^i; i=1,2,\ldots,N'^{}_S\} \cup \{x_B^j;j=1,2,\ldots,N'^{}_B\}, \nonumber \\
                          \\
T^{1\textrm{st}}_{} &=& X^{2}_S \cup \cdots \cup X^{M}_S \cup X^{2}_B \cup \cdots \cup X^{M}_B \nonumber \\
                          &=& \{x^l_S;l=N'^{}_S+1,N'^{}_S+2,\ldots,{N_S^{}}\} \nonumber \\
                          && ~ \cup ~ \{x^m_B;m=N'^{}_B + 1,N'^{}_B + 2,\ldots,{N_B^{}}\}, \label{eq_1st_tra}
\end{eqnarray}
where ${N'^{}_S} = N_S^{}/M$ and ${N'^{}_B} = N_B^{}/M$.
Then, with same manner, other $M-1$ pairs can be configured as
\begin{eqnarray}
E^{2\textrm{nd}}_{} &=& X^{2}_S \cup X^{2}_B \phantom{\cup \cdots \cup X^{M}_S \cup X^{1}_B \cup X^{3}_B \cup \cdots \cup X^{M}_B} \nonumber \\
                           &=& \{x^i_S;i=N'^{}_S+1,\ldots,{2N'^{}_S}\} \nonumber \\
                           && ~ \cup ~ \{x^j_B;j=N'^{}_B+1,\ldots,{2N'^{}_B} \},\\
T^{2\textrm{nd}}_{} &=& X^{1}_S \cup X^{3}_S \cup \cdots \cup X^{M}_S \cup X^{1}_B \cup X^{3}_B \cup \cdots \cup X^{M}_B \nonumber \\
                           &=& \{x^l_S;l=1,\ldots,{N'^{}_S}\} \nonumber \\
                           && ~ \cup ~ \{x^l_S;l=2N'^{}_S+1,\ldots,{N_S^{}}\} \nonumber \\
			  && ~ \cup ~ \{x^m_B;m=1,\ldots,{N'^{}_B}\} \nonumber \\
			  && ~ \cup ~ \{x^m_B;m=2N'^{}_B+1,\ldots,{N_B^{}}\},\label{eq_2nd_tra}
\end{eqnarray}
\begin{eqnarray}
			  \vdots \nonumber
\end{eqnarray}
\begin{eqnarray}
E^{M\textrm{th}}_{} &=& X^{M}_S \cup X^{M}_B \nonumber \\
                           &=& \{x^i_S;i=(M-1)N'^{}_S+1,\ldots,{N^{}_S}\} \nonumber \\
                           && ~ \cup ~ \{x^j_B;j=(M-1)N'^{}_B+1,\ldots,{N^{}_B}\},\\
T^{M\textrm{th}}_{} &=& X^{1}_S \cup \cdots \cup X^{M-1}_S \cup X^{1}_B \cup \cdots \cup X^{M-1}_B \nonumber \\
                           &=& \{x^l_S;l=1,\ldots,{(M-1)N'^{}_S}\} \nonumber \\
                           && ~ \cup ~ \{x^m_B;m=1,\ldots,{(M-1)N'^{}_B}\}. \label{eq_Mth_tra}		
\end{eqnarray}
Note that, from above relations, one can easily see that overlaps in configured samples between different training sets are allowed while there are no overlaps between each of evaluation sets. By repeating the similar paring for other sets we now have $M$ pairs of [$T^k_{}$, $E^k_{}$] which are having different evaluation samples to each others. In this work, we prepare 10 pairs of round-robined samples with setting $M=10$. 


Firstly we train ANN with the samples in a training set $T^k_{}$ and then evaluate $x_S^i$s and $x_B^j$s in a paired evaluation set $E^k_{}$. In the training process, ANN recursively finds an optimal connection weight between each node by the iRPROP algorithm. In practice, we set the tolerance of error between the originally given ranks as either 1 or 0 for $x_S^i$ or $x_B^j$, respectively, and the value of output node to be $10^{-3}$. When the training process is finished, the resulted connection weights are saved for the evaluation process. Then, evaluation samples $x_S^i$ and $x_B^j$ get their ranks based on the trained result. Finally, from the evaluation, ANN scores a rank between 0 and 1 for each evaluation sample. 

\section*{References}
\bibliography{cbcgrbmvcbib}
\bibliographystyle{jphysicsB}

\end{document}